# Enhanced Luminous Transmission and Solar Modulation in Thermochromic VO2 Aerogel-Like Films via Remote Plasma Deposition

Jose M. Obrero-Pérez,[a] Gloria Moreno-Martínez,[a] Teresa C. Rojas,[b] Francisco J. Ferrer,[c,d] Francisco G. Moscoso,[e] Lidia Contreras-Bernal,[a,f] Javier Castillo-Seoane,[a] Fernando Núñez-Gálvez,[a] Francisco J. Aparicio,[a] Ana Borras,[a] Juan R. Sánchez-Valencia,[a*] Angel Barranco[a*]

a) Nanotechnology on Surfaces and Plasma Laboratory, Materials Science Institute of Seville (CSIC-US), C/ Américo Vespucio 49, 41092, Seville, Spain.

b) Tribology and Surface Protection Group. Materials Science Institute of Seville (CSIC-US), C/ Américo Vespucio 49, 41092, Seville, Spain.

c) Centro Nacional de Aceleradores (Universidad de Sevilla, CSIC, Junta de Andalucía). Avda. Tomas Alba Edison 7, 4092, Sevilla

d) Departamento de Física Atómica, Molecular y Nuclear, Universidad de Sevilla, Aptdo 1065, 41012 Sevilla, Spain

e) Center for Nanoscience and Sustainable Technologies (CNATS), Departamento de Sistemas Físicos, Químicos y Naturales, Universidad Pablo de Olavide, Ctra. Utrera km. 1, Sevilla 41013, Spain.

f) Departamento de Ingeniería y Ciencia de los Materiales y el Transporte, EPS-Universidad de Sevilla, c/Virgen de África 7, 41011, Sevilla, Spain.

Corresponding author e-mail: angel.barranco@csic.es; jrsanchez@icmse.csic.es

**Abstract**

Vanadium dioxide ($VO_2$) is a thermochromic material that undergoes a phase transition from a monoclinic semiconducting state to a rutile metallic state at 68 °C, a temperature close to room temperature. This property makes $VO_2$ particularly valuable in applications such as optical and electrical switches, data storage, neuromorphic computing, and remarkably dynamic smart windows for solar radiation control. $VO_2$ typically needs to be synthesized for these applications as nanostructured thin films. Over the past few decades, significant efforts have been made to control the thermochromic properties of $VO_2$ through crystal structure tuning, doping, and the development of $VO_2$ nanocomposites. Additionally, introducing nano- and mesoporosity has been shown to enhance the optical properties of thermochromic $VO_2$ films. This study presents a methodology for producing highly porous, aerogel-like $V_2O_5$ films, which can be thermally processed to form aerogel-like $VO_2$ films. This process is based on sequential plasma polymerization and plasma etching to produce aerogel-like $V_2O_5$ films that are annealed to yield ultraporous nanocrystalline $VO_2$ films. The sacrificial vanadium-containing plasma polymers are obtained by remote plasma-assisted vacuum deposition (RPAVD) using vanadyl porphyrin as a precursor and Ar as plasma gas. Additional reference compact films $VO_2$ films are obtained by a direct RPAVD process using the same precursor and oxygen plasmas in combination with thermal annealing. The aerogel-like $VO_2$ films show exceptional thermochromic performance with luminous transmittances higher than 54%, solar modulation up to 18.8%, and IR modulation up to 35.5%. The presented plasma methodology is versatile, allowing both the synthesis of $VO_2$ plasmonic structures to enhance the thermochromic response and the encapsulation of films to improve their stability in air dramatically. Additionally, this solvent-free synthetic method is fully compatible with doping procedures, scalable, and holds great potential for designing and optimizing smart window coatings.

## 1.- Introduction

The escalating demand for high energy, driven primarily by innovative technologies, has led to a significant increase in carbon dioxide emissions, directly contributing to global warming. This demand constitutes approximately 30-40% of the overall energy consumption in conventional households.[1] To mitigate this surge, the development of materials capable of reducing environmental impact while enhancing indoor comfort, particularly concerning heating/cooling, ventilation, and lighting, has become imperative. In this context, materials exhibiting thermochromic properties present a promising solution. These materials can transmit the entire solar spectrum at low temperatures and selectively reflect infrared radiation at higher temperatures. In recent decades, vanadium (IV) oxide has emerged as a promising candidate for smart windows due to its reversible metal-insulator transition (MIT) at a relatively low transition temperature ($T_t$) of 68 ºC (341K) for bulk material.[2] Below this threshold, VO$_2$ maintains a monoclinic lattice (P2$_1$/c, VO$_2$(M)) that switches to rutile (P4$_2$/mnm, VO$_2$(R)) above this temperature. This phase transition is accompanied by a substantial change in the optical and electrical properties, which are reversible by cooling.[3] The VO$_2$(M) phase is semiconducting and transparent to infrared solar radiation, while VO$_2$(R) is metallic and reflects solar radiation. In the rutile phase, VO$_6$ octahedrons share vertices to form a compact hexagonal lattice where the distance between the vanadium atoms at the centers of each octahedron is fixed ($\sim$ 2.85 Å). During the transition phase, vanadium atoms move alternately along the V-V direction, forming short and long V-V distances (3.12 Å and 2.65 Å), distorting the zigzag VO$_6$ octahedrons and inducing semiconductor behavior.[3–5] This phenomenon can be elucidated through Molecular Orbital theory, wherein bonding ($\sigma$, $\pi$) and antibonding ($\sigma$*, $\pi$*) molecular orbitals are formed by the union of V3d and O2s atomic orbitals. In VO$_2$(R), only a d$_{II}$-no bond is formed between adjacent V$^{4+}$ orbitals along the c-axis, causing an overlap between dII and $\pi$*, and the Fermi level falls, giving it metallic characteristics. Nevertheless, two V-V distances in VO$_2$(M) result in dII-bonding and dII-antibonding components, raising the Fermi level and generating an insulating phase.[6–8]

In the field of smart windows, key limitations hinder the performance of thermochromic VO$_2$-based materials. For optimal thermochromic behavior, the main requirements include an appropriate transition temperature (T$_t$), high luminous transmittance (T$_{lum}$), a significant modulation of solar transmittance ($\Delta$T$_{sol}$), and good stability against oxidation.[1] To effectively implement VO$_2$ coatings on conventional glass for smart window applications, the material must fulfill the following criteria: (i) Tt close to ambient temperature ($\sim$ 25 ºC);[9,10] (ii) T$_{lum}$ > 60%;[11–13] (iii) $\Delta$T$_{sol}$ > 10%;[14–16] (iv) excellent near-infrared (NIR) transmittance modulation capacity ($\Delta$T$_{IR}$) > 10%[17,18] and (v) long-term stability.[19] Numerous strategies have been explored to lower the transition temperature while improving the thermochromic and solar energy efficiency of a VO$_2$ film.[20] However, the improvement of $\Delta T_{sol}$ typically occurs at the expense of T$_{lum}$, which implies a complex challenge when optimizing the synthesis of these films. The film microstructure modification appears to be the key to addressing this challenge. In recent years, a certain balance between $\Delta T_{sol}$ and T$_{lum}$ has been found to improve thermochromic efficiency by modifying part of the VO$_2$ structure, such as crystal morphology,[21] particle size,[22] and porosity.[23,24] Other common strategies to decrease T$_t$ include doping[12,25–28] and co-doping[29–31] the VO$_2$ with different elements. However, balancing the transition temperature with optical properties poses an even more significant challenge. Nonetheless, previous studies have demonstrated that, in comparison to other approaches, introducing increased porosity as a secondary component into VO$_2$ films offers a viable alternative for enhancing $\Delta T_{sol}$ without compromising T$_{lum}$ and $\Delta T_{IR}$,[20,24,32] being this strategy also beneficial for lowering the phase transition temperature.[33]

We have recently reported a cyclic plasma-based process that enables the synthesis of aerogel-like, conformal TiO$_2$ thin films with low density and ultra-high porosity values characteristic of aerogel materials[34]. In this work, we extend this approach to fabricating aerogel-like thermochromic VO$_2$ films. Thus, the aerogel-like VO$_2$ film deposition combines the plasma polymerization by remote plasma-assisted vacuum deposition (RPAVD)[35–41] of a V-containing precursor and remote soft plasma etching (SPE) followed by a reduction thermal treatment to get the appropriate phase and stoichiometry. Our research delves into the impact of film thickness and porosity on their thermochromic properties, including T$_{lum}$, $\Delta T_{sol}$, and $\Delta T_{IR}$, as well as their transition temperature. We have also explored the encapsulation of the films with a transparent plasma polymer to preserve the stability of the VO$_2$ aerogel-like structure in the face of oxidation. This work presents the fundamental principles of this innovative method for producing robust and high-performance thermochromic $VO_2$-based coatings suitable for smart window applications by industrially scalable procedures.

## 2. Experimental section

### 2.1 Fabrication of aerogel-like $VO_2$

The experimental methodology and setup are fully described in reference[39] applied to the synthesis of aerogel-like $TiO_2$, and it consists of a cyclic procedure combining remote plasma polymerisation (RPAVD) and plasma etching (SPE) to yield porous oxide films. This combined procedure is repeated several times to increase the thickness of the films. The films were deposited on pieces of Si (100) and fused silica slides. The characteristics of each synthetic step are the following:

*a) Deposition of V-containing plasma polymer.* The precursor Vanadyl phthalocyanine (VOPc, dye content >90%) from Sigma-Aldrich was used as received. VOPc plasma polymers were deposited by remote plasma-assisted vacuum deposition (RPAVD). The details of the RPAVD process and a complete description of the experimental setup is reported elsewhere.[36,38] In resume, the plasma polymerization was carried out in an electron cyclotron resonance (ECR) 2.45 GHz microwave (MW) plasma reactor. The sample holder was located in the downstream region of the plasma at ~9 cm from the Ar discharge, with the samples facing down.[39] The precursor was sublimated from a Knudsen cell facing the samples at ~9 cm from the holder position under an Ar plasma of 150 W at $10^{-2}$ mbar. Before deposition, the system was evacuated to a base pressure of $10^{-6}$ mbar. Argon (Ar) gas was introduced using a calibrated mass flow controller. The deposition rate of the growing film was monitored with a quartz crystal microbalance (QCM) positioned near the sample holder. Under these conditions, deposition was performed at room temperature, as measured by an encapsulated thermocouple connected to the sample holder. The VOPc RPAVD plasma polymers were used as sacrificial layers for the synthesis of ultraporous oxides by plasma etching.

**b) Remote soft plasma etching (SPE-O$_2$).** Plasma etching treatment of the VOPc films was carried out in the same reactor, rotating the sample holder to face the plasma, reducing the distance to the plasma by 4 cm, and heating the sample holder to 150 °C. O$_2$/Ar gas (1:1) was introduced as plasma gas. The treatment was carried out at $10^{-3}$ mbar with a plasma power of 300 W. The treatment duration was 60 minutes. These etching conditions (i.e., the long treatment times) were chosen to guarantee complete polymer oxidation to produce the ultraporous V$_2$O$_5$ films.

**c) Development of V$_2$O$_5$ ultra-porous films with controlled thickness.** The RPAVD and SPE steps were repeated several times to control the thickness of the ultra-porous V$_2$O$_5$ film. Hereafter, we name the combination of these two steps a synthetic cycle C$_n$, being *n* the number of repetitions. This cyclic procedure allowed us to incrementally increase the thickness of the oxide films up to several hundreds of nanometers, as shown in the following sections.

**d) Fabrication of reference compact V$_2$O$_5$ films.** A set of V$_2$O$_5$ referred hereafter as compact reference films, were deposited using a modified procedure using the same experimental setup and samples location in the reactor by sublimating VOPc in the presence of a remote reactive remote O$_2$ plasma (RPAVD-O$_2$). The experimental conditions were a pressure of $10^{-3}$ mbar and 300 W of MW power. The RPAVD-O$_2$ was carried out at room temperature. These deposition conditions result in yellowish V$_2$O$_5$ films. A 10-minute after-deposition O$_2$ plasma treatment (SPE) was applied at the same MW power and pressure to ensure complete oxidation of the sample surfaces.

*e) Synthesis of VO₂ films by thermal annealing.* The $V_2O_5$ films produced using the previously described synthetic procedures were thermally annealed to yield stoichiometric $VO_2$ films. The thermal process was carried out in an atmosphere-controlled furnace in an argon environment (70 sccm Ar flow) using a EUROTHERM 2408 temperature controller. A heating ramp of 5 ºC·min⁻¹ was applied. This annealing process was carried out for 30 minutes at 480 ºC to achieve the maximum conversion of V(V) to V(IV) (see next sections). For aerogel-like samples with thicknesses below ~200 nm, the annealing process was shortened to 5 minutes. Afterward, the film was left to cool down back to room temperature

*f) Thin film encapsulation by Adamantane remote plasma polymers.* Adamantane plasma polymers were deposited by RPAVD to encapsulate selected porous $VO_2$ samples in the same plasma reactor. A complete description of this deposition procedure can be found elsewhere[38,41] In resume, adamantane powder (≥99%, Sigma-Aldrich) was sublimated inside the chamber from a heated container at 40 ºC in the presence of an Ar plasma at $2 \cdot 10^{-2}$ mbar and 150W. The deposition was carried out at room temperature.

## 2.2. Characterisation methods

High-resolution field emission scanning electron microscopy (FESEM) images of the samples deposited on silicon wafers were obtained in a Hitachi S4800 field emission microscope, working at 2 kV. Cross-sectional views were obtained by cleaving the Si (100) substrates. Scanning transmission electron microscopy (STEM), high-resolution transmission electron microscopy (HRTEM), and high-angle annular dark field (HAADF)-STEM images were acquired in a Tecnai G2 F30 S-Twin STEM from FEI, equipped with a HAADF detector from Fischione with a 0.16 nm point resolution. For this, portions of each sample type were mechanically scrapped and placed in the microscopy grids. Electron energy loss spectroscopy (EELS) was performed using a QUAMTUM Gatan Imaging Filter (GIF) attached to the microscope. The Electron energy loss spectra (EELS) were collected in Diffraction mode using a camera length of 560 mm with a spectrometer collection angle of 0.7 mrad. To minimize electron beam irradiation and prevent the reduction of the sample, a smaller condenser aperture and a small probe size were employed, reducing the current and electron dose.[42,43] This approach was necessary as vanadium oxides are prone to reduction under high electron beam doses. Under these conditions, the energy resolution system was ~1 eV. After experimental acquisition, the data were processed using the Gatan Digital Micrograph software. Selected area electron diffraction patterns (SAED) were also recorded with low electron dose. $V_2O_5$ and vanadium (IV) acetyl-acetonate powders from Aldrich were used as reference samples of $V^{5+}$ and $V^{4+}$ for the EELS characterization of the samples. These EELS patterns of the references are the same as those reported in the bibliography for $V_2O_5$ and $VO_2$.[42,44]

X-ray diffraction at a grazing incidence (GIXRD) using Cu Kα (50 kV, 1mM) analysis was carried out on a Bruker D8 Discover diffractometer coupled to a 2D detector (Eiger 2R 500K), with an angle of incidence set at 0.5º. The measurement span ranged from 10 to 60º, with increments of 0.02º and a time step of 90 minutes.

Variable angle spectroscopic ellipsometry (VASE) was acquired in a Woollam V-VASE ellipsometer. The optical constants were modeled by fitting the spectra to the Cauchy model.

Rutherford Backscattering Spectroscopy (RBS) and Nuclear Reaction Analysis (NRA) characterizations were performed at the 3 MV Tandem Accelerator of the National Centre of Accelerators (Seville, Spain). RBS measurements were performed with α-particles of 2.0 MeV and a passivated implanted planar silicon (PIPS) detector set at a 165° scattering angle. NRA was used to determine C, N, and O elements in the films from the $^{12}C(d,p)^{13}C$, $^{14}N(d,\alpha_1)^{12}C$ y $^{16}O(d,p^1)^{17}O$ nuclear reactions using deuterons of 1.0, 1.4 y 0.9 MeV, respectively. The spectra were obtained using a particle detector set at 150º collection angle and a 13 μm thick Mylar filter to stop the backscattered particles. NRA and RBS spectra were simulated and fitted using the SIMNRA 6.0 code.[45] The film densities were determined from the combined RBS and NRA analyses of the content per square centimeter and thickness values obtained from cross-sectional FESEM micrographs of the films.[46]

XPS characterizations were performed in a Phoibos 100 DLD X-ray spectrometer from SPECS. Before analysis, samples underwent pre-treatment via Ar ion bombardment ($10^{-6}$ mbar, 5 kV, 5 minutes). The spectra were collected in the pass energy constant mode at 50 eV using an Mg Kα source. C1s signal at 284.8 eV was used to calibrate the spectra's binding energy (BE). The assignment of the BE to the different elements in the spectra corresponds to the data in references.[47–50]

Optical transmittance properties of the samples deposited on fused silica substrates have been analyzed in the 200-2500 nm wavelength range recorded in a PerkinElmer Lambda 750 S UV–vis–NIR spectrophotometer. In-situ measurements were carried out with a homemade device to characterize hysteresis curves. It consisted of two ceramic heater-resistor holders with a concentric pierced 5 mm diameter hole at the center, allowing light transmission, and was used to hold the samples during the spectra acquisitions. The plates were connected to an ISOTECH IPS-405 direct current (DC) power source, allowing uniform sample heating from both sides. Additionally, one of the heating plates was equipped with an encapsulated thermocouple placed close to the samples and connected to a temperature reader. The transmittance versus temperature thermochromic hysteresis loops were obtained at a wavelength of 2000 nm in variable steps to temperature, ranging from room temperature (RT) to 110 ºC. The current was gradually applied until the target temperature was reached. After achieving each desired temperature, an equilibration period was observed to attain thermal equilibrium and compensate for any minor temperature fluctuations. The same device was also used for in-situ Raman spectroscopy characterizations using a Horiba Jobin-Yvon LabRAM spectrometer equipped with a confocal microscope with a 50x objective and a green laser of 532 nm wavelength. The spectral resolution for this configuration was $\sim 1.7$ cm$^{-1}$. No polarisation was applied during the experiments. Low laser powers were utilized to prevent local heating that could lead to oxidation of the VO$_2$ films and unintended phase transitions.

### 3. Results and Discussion

### 3.1 Synthesis of V$_2$O$_5$ films

**Figure 1 a)** resumes the procedure for synthesizing aerogel-like V$_2$O$_5$ thin films consisting of one or several cycles of combined VOPc RPAVD of vanadium-containing plasma polymers and remote soft plasma etching. The VOPc plasma polymer films present an intense light absorption related to the VOPc bands, as shown in **Figure S1a)**, and present a continuous and homogeneous microstructure (**Figure S1b)**), with stoichiometry similar to the precursor molecule. In each cycle, the VOPc plasma polymer acts as a sacrificial layer for the ultraporous V$_2$O$_5$ films by plasma etching. Thus, the V-containing plasma polymer conformally coats the substrate surface (C$_1$ cycle) or the previously deposited porous oxide films (C$_2$-C$_n$ cycles) before being fully oxidized (see the schematic in **Figure 1a)**). The oxygen plasma etches the organic part of the polymer films, generating low-dense inorganic V$_2$O$_5$ films by the oxidation of the V$^{4+}$ cations of the VOPc films. Simultaneously, carbon and nitrogen elements of the VOPc polymer form oxygenated volatile species that are pumped out of the chamber. This deposition procedure is analogous to the recently reported method for synthesizing aerogel-like partly crystalline TiO₂ optical films using a Ti(IV) phthalocyanine precursor.[34]

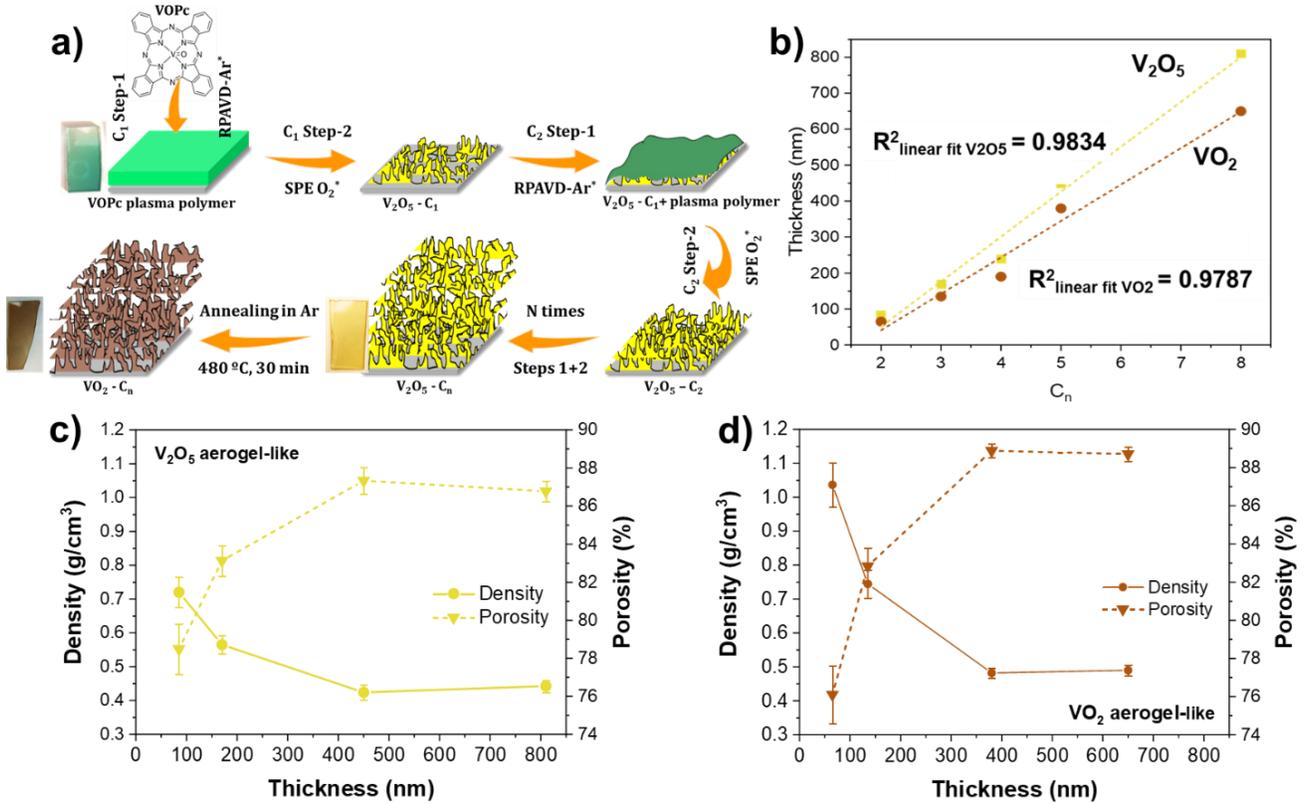

**Figure 1.** (a) Scheme of the cyclic procedure for fabricating aerogel V₂O₅ and VO₂ films. (b) Relation between film thickness and the number of cycles for aerogel-like V₂O₅ films and the corresponding VO₂ aerogel-like after annealing. The V₂O₅ films were synthesized in each cycle using sacrificial V-containing plasma polymers of ~200 nm. (c-d) Density and porosity values as a function of film thickness for V₂O₅ (c) and VO₂ aerogel-like films (d), these latter after the thermal annealing of films in (c).

The film thickness exhibits quasi-linear growth with the number of deposition and etching cycles, as shown in **Figure 1b)**. The thickness reduction observed after Ar thermal annealing, which transforms the $V_2O_5$ to $VO_2$, is consistent with the expected change in bulk densities between these two oxides ($\rho_{V2O5}$=3.36 g·cm$^{-3}$; $\rho_{VO2}$=4.34 g·cm$^{-3}$).[51] **Figure 1c)** shows the evolution of $V_2O_5$ film density deposited by RPAVD-Ar/SPE procedure as a function of the layer thickness. The Figure also shows the corresponding porosity values calculated using the relationship $\Gamma(\%)$=100·(1-$\rho$/$\rho_r$), taking the density of reference crystalline $V_2O_5$ as $\rho_r$= 3.36 g·cm$^{-3}$.[52] The first RPAVD-Ar/SPE cycle yielded films with a density as low as 0.72 g·cm$^{-3}$ and a porosity of 78.5 %. As the layer thickness increases, the density decreases dramatically, reaching a value of 0.42 g·cm$^{-3}$ at 450 nm of thickness, which corresponds to a film porosity as high as 87.3 %. Further increases in thickness led to slight changes in these values, with the density increasing to 0.44 g·cm$^{-3}$ and the porosity increasing to 86.8% at a thickness of 810 nm. Thus, the density and porosity values obtained are in the range of those reported for $V_2O_5$ aerogels obtained by supercritical drying methods[53] and $V_2O_5$ aerogel thin films deposited by sol-gel.[32,54] For this reason, we refer to them hereafter as aerogel-like $V_2O_5$ films.

**Figure 2a)** showcases the planar and cross-sectional films of $V_2O_5$ aerogel-like thin films deposited on Si (100) that correspond to the synthetic cycles C1-C3 and C5 using VOPc sacrificial layers of ~200 nm per cycle. From the first cycle, the films present a unique, low-dense sponge-like microstructure of interconnected percolated oxide structures surrounding a homogeneous distribution of quasi-circular pores. The overall percentage of voids in the structure is very high. Note that all the solid oxide structures in the micrographs are highly porous and fully interconnected. **Figure 2c)** also shows that such a unique porous structure is interconnected. As the number of cycles increases, the size of the porous increases, retaining the sponge-like structure and forming an expansive network where empty spaces prevail. This open porous low-density structure is congruent with the measured density values shown in **Figure 1 c)**. Notably, compared with aerogel-like TiO$_2$ films prepared by a similar procedure,[34] the fully interconnected foam-like structure of the vanadium oxide films is distinctive.

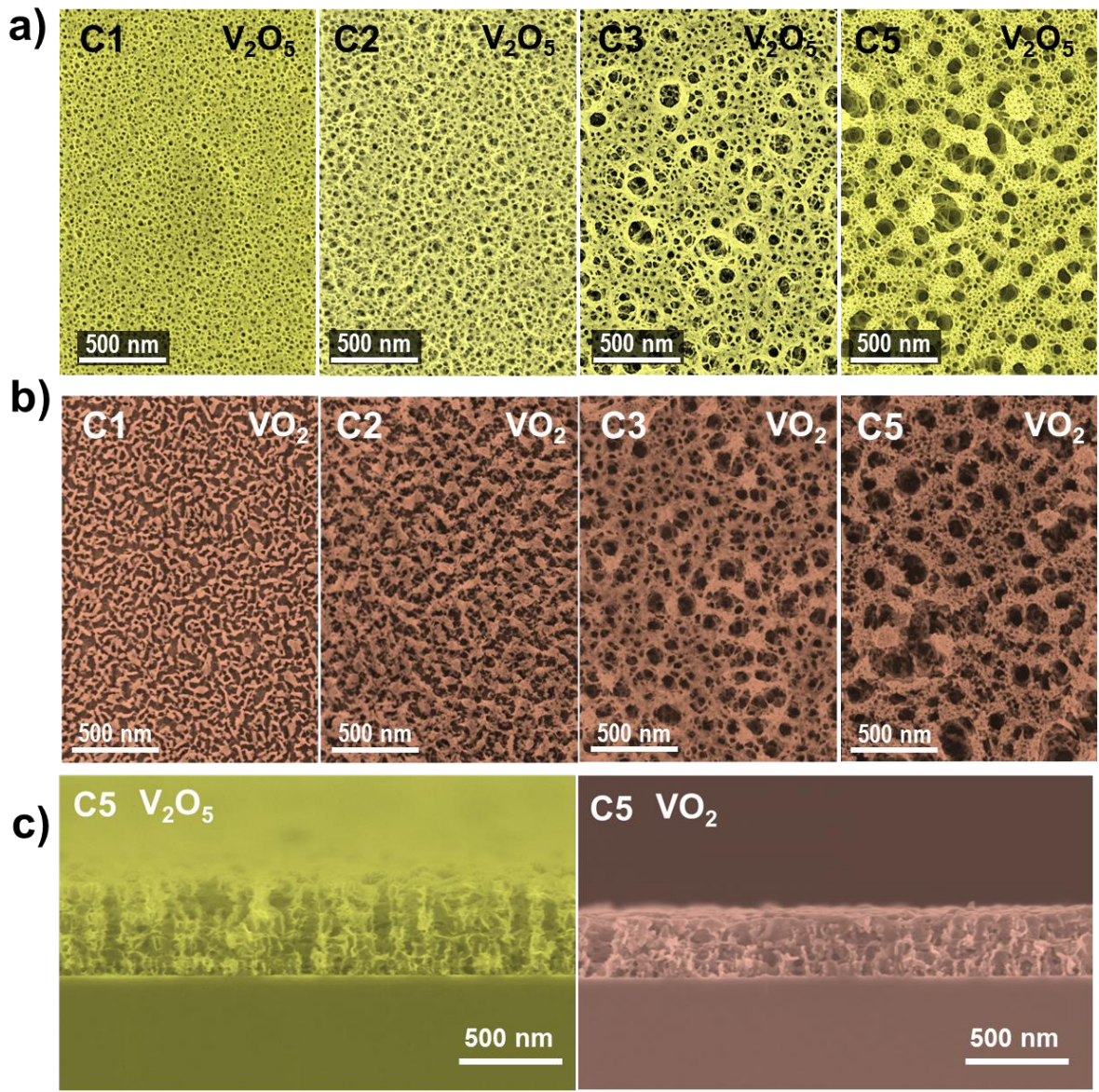

**Figure 2.** (a) Normal view FESEM micrographs of $V_2O_5$ aerogel-like films corresponding to cycles C1 to C5. (b) Normal view FESEM micrographs of $VO_2$ obtained after annealing. (c) Cross-sectional SEM micrographs of a C5 aerogel-like thin film, showing before (left) and after (right) annealing.

**Figure 3a)** shows FESEM micrographs corresponding to $V_2O_5$ films synthesized by RPAVD-$O_2$ with 40 and 110 nm thicknesses, respectively. The samples display a homogenous and featureless microstructure in the thickness range studied. Samples fabricated by this technique are limited to lower thicknesses than aerogel-type samples due to their characteristic optical properties (i.e., high absorptance in the visible and NIR range, as will be shown in the following sections). These films will be used as a reference for properly comparing the optical properties of the high porosity layers synthesized by the cyclic plasma polymerization and plasma etching procedure.

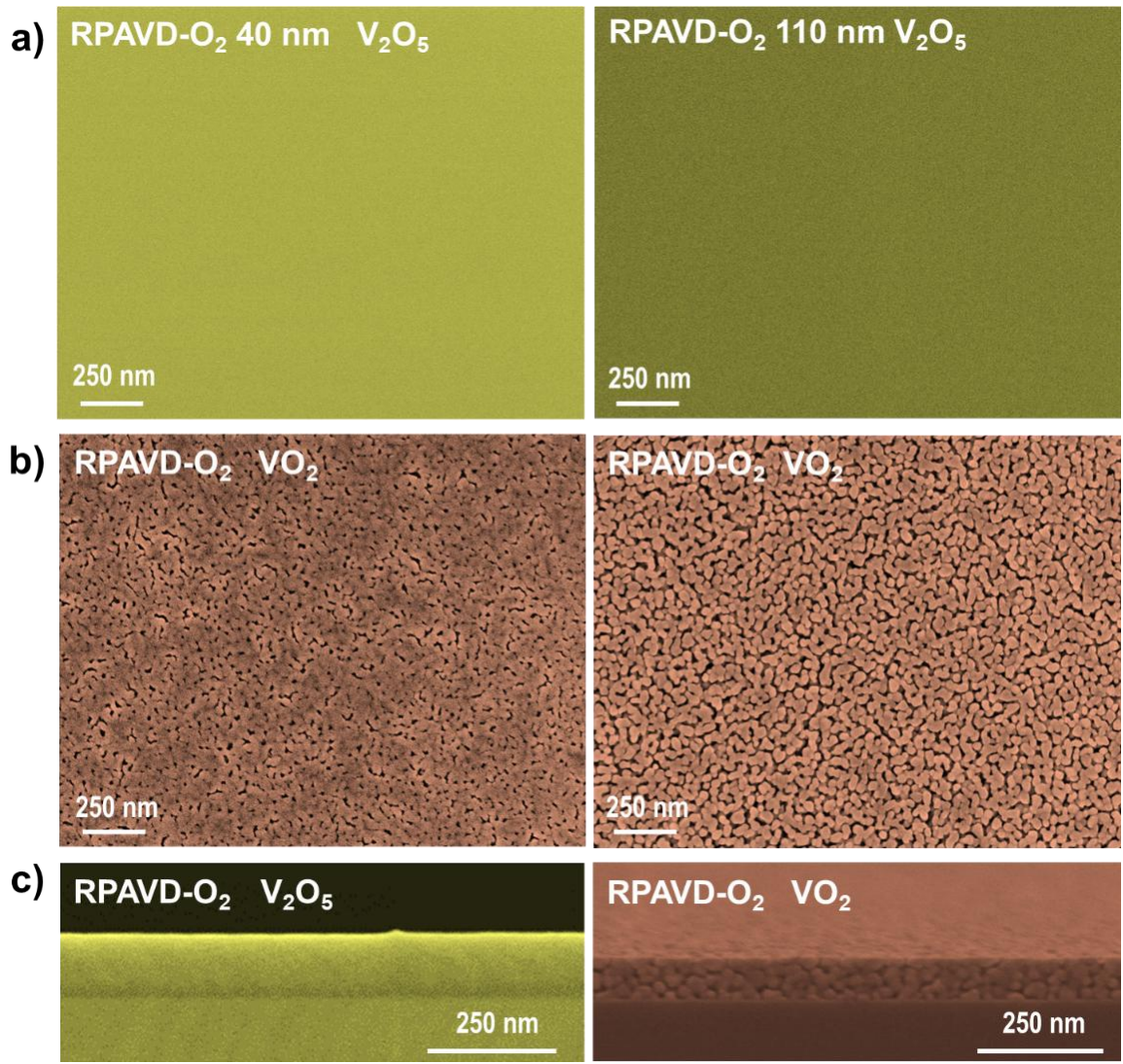

**Figure 3.** (a) FESEM images of two $V_2O_5$ films of 40 nm (left) and 110 nm (right) films deposited by RPAVD-$O_2$. The inset shows a cross-sectional view (b) FESEM images of the thin films in (a) after thermal annealing. (c) Cross-sectional FESEM micrographs of the RPAVD-$O_2$ 110 nm $V_2O_5$ and annealed $VO_2$ films in (a) and (b), respectively.

### 3.2 Synthesis of VO₂ films by annealing

The following experimental step after the film synthesis consisted of determining the conditions for the thermal reduction of the plasma-deposited $V_2O_5$ films to get thermochromic $VO_2$ films. The criterion selected to determine the optimum annealing treatment conditions was to maximize the thermochromic response of the samples, that is, to maximize the difference between the light transmission in the infrared region below and above the transition temperature. Note that a complete analysis of the thermochromic response of the films will be presented in the next section.

**Figure 4a)** shows the normalized transmittance variation between room temperature and 90 ºC (to be sure that the temperature is well below and above the $T_t$), at $\lambda = 2000$ nm, of a plasma-deposited $V_2O_5$ film subjected to annealing at different temperatures for 0.5 hours (under Ar atmosphere). It is observed that below 400 ºC, the transformation from $V_2O_5$ to $VO_2$ (M) is negligible. Above this temperature, the formation of $VO_2$ (M) crystals becomes evident. The transmittance variation (note that this ∆T is normalized) increases as the annealing temperature rises above 400 ºC, reaching a peak at 480 ºC. At temperatures exceeding 480 ºC, the transmittance variation decreases, very likely due to the transformation of $VO_2$ (M) into more reduced species.

**Figure 4 b)** illustrates the evolution of the bandgap for a $V_2O_5$ aerogel-like film with temperature calculated using the Tauc-Plot method[55] as a function of the annealing temperature. As the temperature rises, the bandgap value decreases gradually, ascribed to the transformation from $V_2O_5$ to $VO_2$ (the latter having a bandgap between 0.6 and 1.0 eV).[56,57] The band gap reaches its minimum at 480 ºC, with an average value of 0.6 eV. The bandgap evolution confirms that temperature is optimal for producing $VO_2$ thermochromic films. The evolution is more gradual than the sharp increase in transmittance shown in **Figure 4a)**, indicating that the microstructural and chemical transformation in the films starts at mild temperatures before impacting the thermochromic transition temperature.

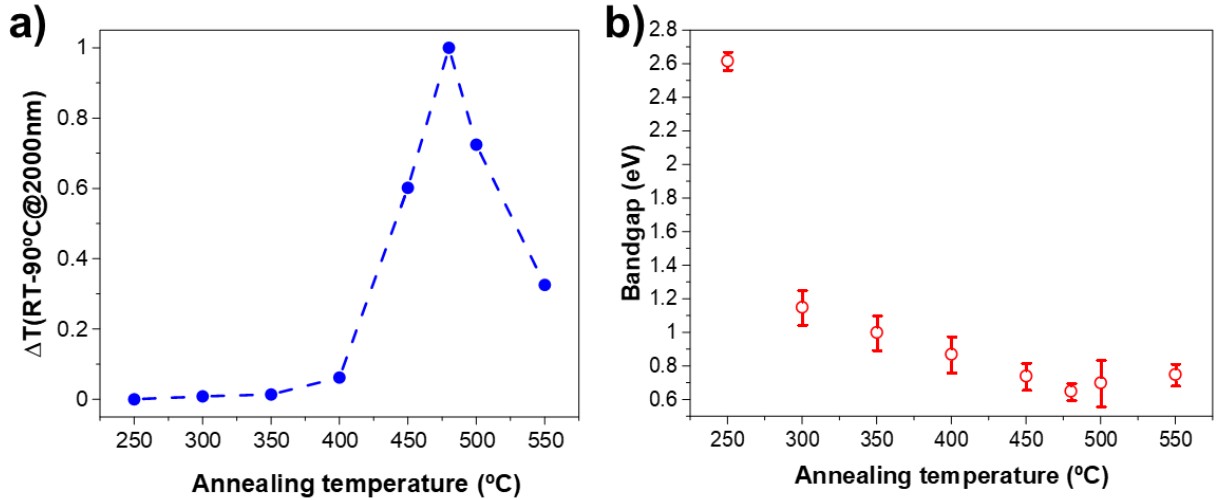

**Figure 4**. (a) Normalised transmittance variation at $\lambda = 2000$ nm between room temperature and 90 ºC for as-deposited $V_2O_5$ aerogel-like samples annealed at different temperatures. The annealing time at each temperature was 0.5 h. (b) Calculated optical bandgap energy by the Tauc-plot method of the $V_2O_5$ aerogel films in (a) as a function of the annealing temperature.

As mentioned before, when the $V_2O_5$ aerogel samples undergo the annealing process, there is an approximate 20% reduction in the thickness of the films, regardless of the initial value (see **Figure 1b**). **Figure 2b)**, shows the FESEM images of the films after the thermal annealing. Compared with the corresponding $V_2O_5$ aerogel-like films, a modification in the microstructure of the newly formed $VO_2$ films is observed after the annealing, leading to more aggregated oxide structures. In the thinner films ($C_1$ and $C_2$), the aggregation of the solid part results in a higher volume of macroporous structures. However, in the case of the thicker films ($C_3$ and $C_5$), the aggregation of the oxide structures is also observed, but the films present a pore distribution and oxide structures percolation that resembles that of the starting $V_2O_5$ films. **Figure 1d)** shows the evolution of film density and porosity after the annealing process. The porosity values are determined considering a reference density for $VO_2$ of 4.34 $g \cdot cm^{-3}$.[52]

The density values are, in all cases, higher than those of the corresponding $V_2O_5$ films (**Figure 1c)**). Thus, from an initial value of 0.72 $g \cdot cm^{-3}$ for the thinner film, 1 $g \cdot cm^{-3}$ is reached for the annealed film. The density of the $VO_2$ films decreases as the thickness of the films increases while remaining, in all cases, above the values of the starting $V_2O_5$ oxide films. The values for the thicker layers are 0.48 and 0.49 $g \cdot cm^{-3}$ for the 380 and 650 nm films, respectively. The porosity values are relatively high, reaching 82.8 % for the 135 nm thick sample and higher than 88.5 % for the thicker layer. These values are in the range of aerogel materials. Note that the higher porosity values of the $VO_2$ films after annealing depend on the density value of the $VO_2$ reference used in the calculation, which is higher than that of the $V_2O_5$ reference.

Contrarily, when the reference RPAVD-$O_2$ $V_2O_5$ samples are annealed to get $VO_2$ thin films (**Figure 3b**), the microstructure of the films undergoes a significant transformation. The resulting structure manifests as a rugged surface morphology comprising minute irregular granules ranging in size from 35 to 70 nm. This aggregated structure is more pronounced as the thickness increases. The cross-sectional view in **Figure 3c)** shows a similar aggregated microstructure in the film cross-section. Besides, an open porous structure between the solid aggregates is also patent in the films, being more pronounced for the thickest film.

### 3.3 Thin film crystallinity and surface composition

**Figure 5a)** shows TEM, HAADF-STEM, SAED pattern, and HRTEM micrographs of as-deposited $V_2O_5$ aerogel-like film. These images reveal an ultraporous structure consisting of roughly spherical cavities with wall thicknesses ranging from 5 to 10 nm. The sample exhibits both amorphous and nanocrystalline phases, as indicated by the broad and diffuse halo and the diffraction rings observed in the SAED pattern and by the lattices fingers and amorphous regions visible in the HRTEM image. After annealing at 480°C (**Figure 5b**), the sample still exhibits a high degree of porosity, as observed in the representative TEM micrograph, which also reveals crystals ranging from 10 to 20 nm in size. The SAED pattern (inset of **Figure 5b**) confirms the formation of a polycrystalline structure.

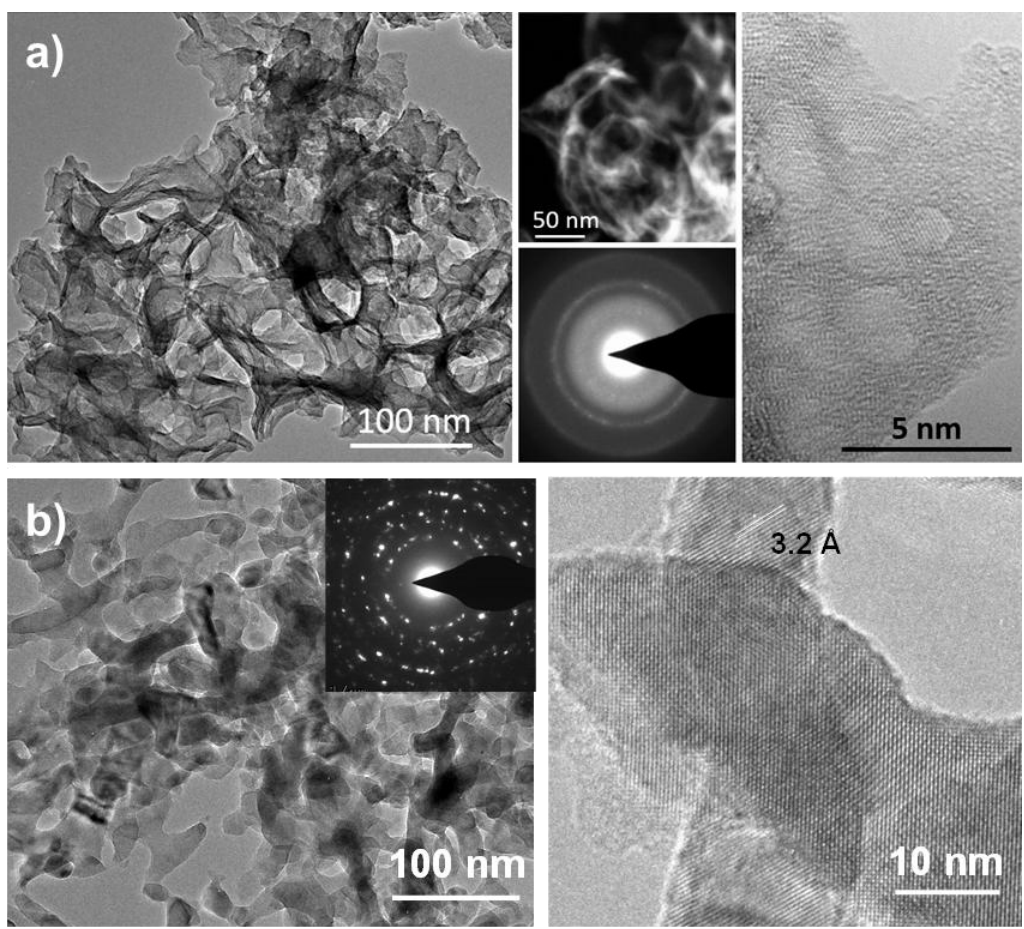

**Figure 5.** (a) TEM micrograph (left), HAADF-STEM image and SAED pattern (center), and HRTEM micrograph (right) of an Aerogel-like $V_2O_5$ film. (b) TEM image, SAED pattern (left), and HRTEM micrograph (right)of a VO2 film after annealing.

The energy loss near-edge structure (ELNES) of the oxygen K-edge in the EELS spectra in vanadium oxide provides a valuable insight into the local density of states (LDOS) at the oxygen site, serving as a reliable indicator of the material's oxidation states.[44,58] Changes in the vanadium valence state lead to a relative chemical shift between the oxygen K-edge and vanadium L3-edge. The ELNES features and the relative chemical shifts for standard $V_2O_5$($V^{5+}$) and $VO_2$($V^{4+}$), have been well documented in the bibliography.[44] **Figure 6** compares the EELS spectra from an aerogel-like $VO_2$ sample with those of $V^{5+}$ and $V^{4+}$ references (see experimental section). The aerogel-like film in the figure has a distance between V-$L_3$ and O-K $e_g$ peaks ($\Delta E$) and relative intensities of the O-K $t_{2g}$ and $e_g$ peaks that are similar to the $VO_2$ reference values,[44,58] which is consistent with the formation of $VO_2$ after the annealing. Note that the $V_2O_5$ reference presents 1 eV less in the value of $\Delta E$ (13 vs 14 eV), as reported in literature.[44,58]

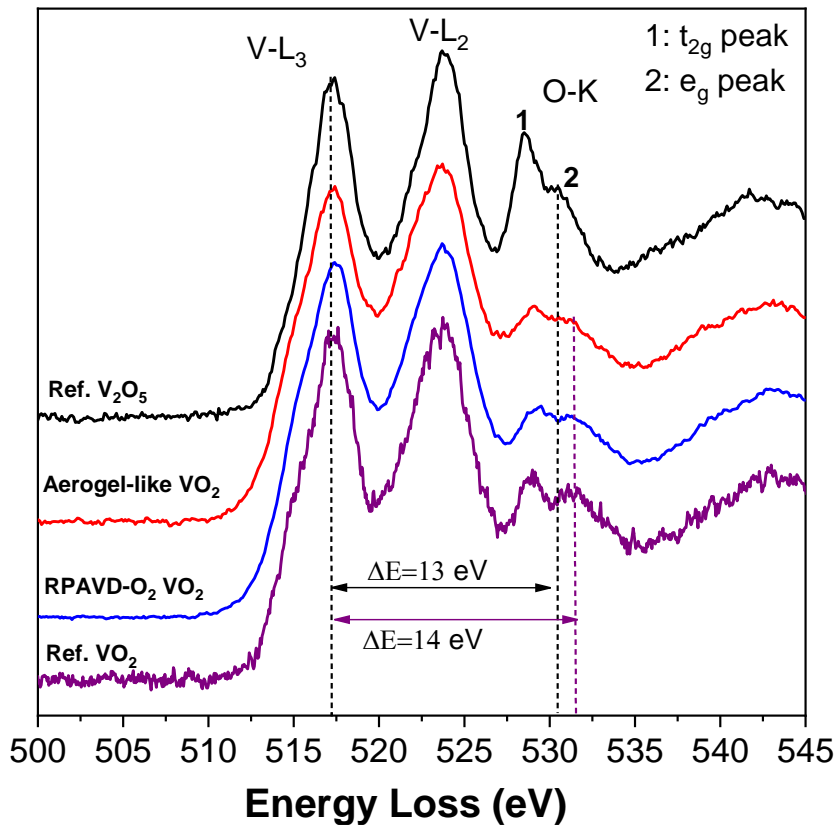

**Figure 6.** Energy loss near edge structure (ELNES) of aerogel-like $VO_2$ and RPAVD-$O_2$ samples and $VO_2$ and $V_2O_5$ reference spectra. The energy separation ($\Delta E$) between the O K-edge ($e_{g.,}$ peak) and the V-$L_3$ edge for the two reference samples is indicated in the figure.

Figure 7a shows SAED patterns of the aerogel-like $VO_2$ film. The left pattern corresponds to the as-prepared state, while the central and right patterns were obtained after progressive heating under the electron beam. The ring spacings measured in the initial pattern (left) correspond to the monoclinic $VO_2$(M) phase, consistent with the XRD analysis (ICSD No. 60767). However, the measured values are slightly larger, suggesting an increased lattice parameter land indicating lattice strain. Specifically, the distance for the first diffraction ring is measured at 3.5 Å, compared to the theoretical 3.2 Å for the (0-11) plane (ICSD No. 74705). Upon beam-induce heating, this spacing decreases to 3.4 Å (central pattern) and finally to 3.2 Å after prolonged exposure (right), indicating a transition. The decreased distances in the right pattern of **Figure 7a)**, can be attributed to the rutile $VO_2$ (R) phase (ICSD No. 66665). **Figure 7b)** displays the evolution of the diffraction patterns with increasing electron beam irradiation time and heating of the sample.

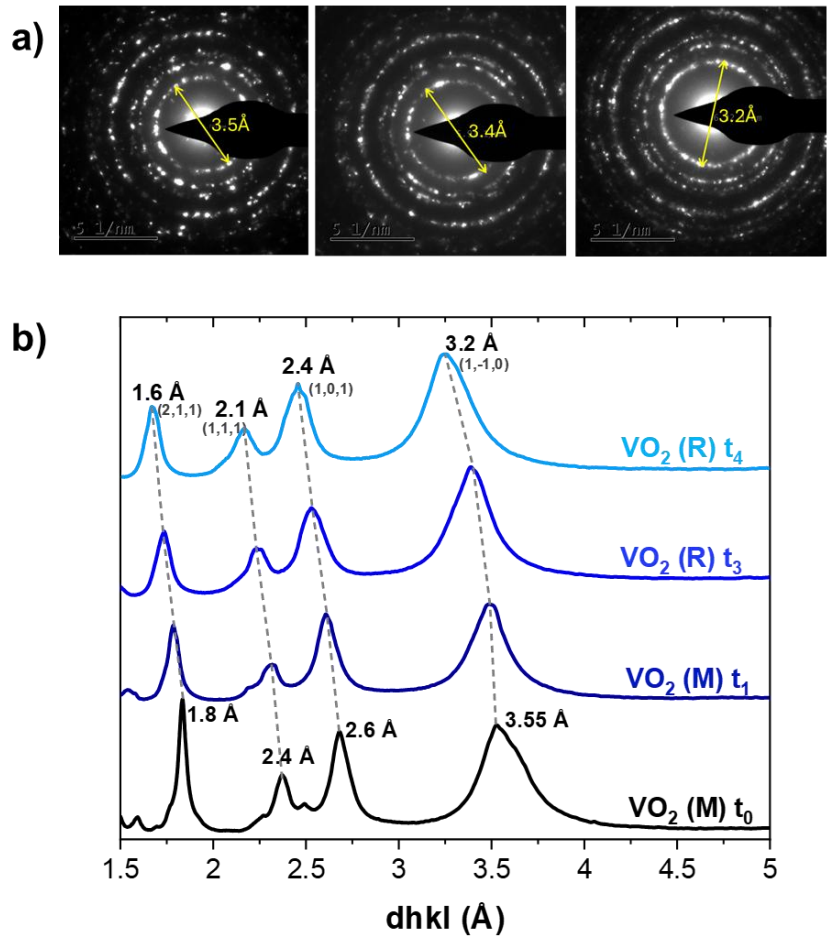

**Figure 7.** (a) SAED of the aerogel-like VO$_2$ film heating by electrons by increasing time. The diffraction patterns correspond to the initial state (left) and after heating under the electron beam (centre) and (right). (b) Plot of the intensity of the diffraction rings in (a) for increasing periods from t$_0$ to t$_4$.

TEM analysis of an RPAVD-O$_2$ V$_2$O$_5$ sample (**Figure 8a**)) reveals that the film comprises packed oxide nanocrystals ranging from 20 to 60 nm in size. The SAED pattern shows rings characteristic of polycrystalline samples with no evidence of an amorphous phase. The diffraction rings can be assigned to V$_2$O$_5$. After annealing (**Figure 8b**)), the sample is still composed of nanocrystals with a slight increase in crystal size to approximately 30-90 nm. The rings´ spacings in the figure correspond to $VO_2$, indicating a phase transformation. Furthermore, the ELNES spectrum of this film (**Figure 6**) confirms that, after annealing, the film predominantly contains $V^{4+}$ species.

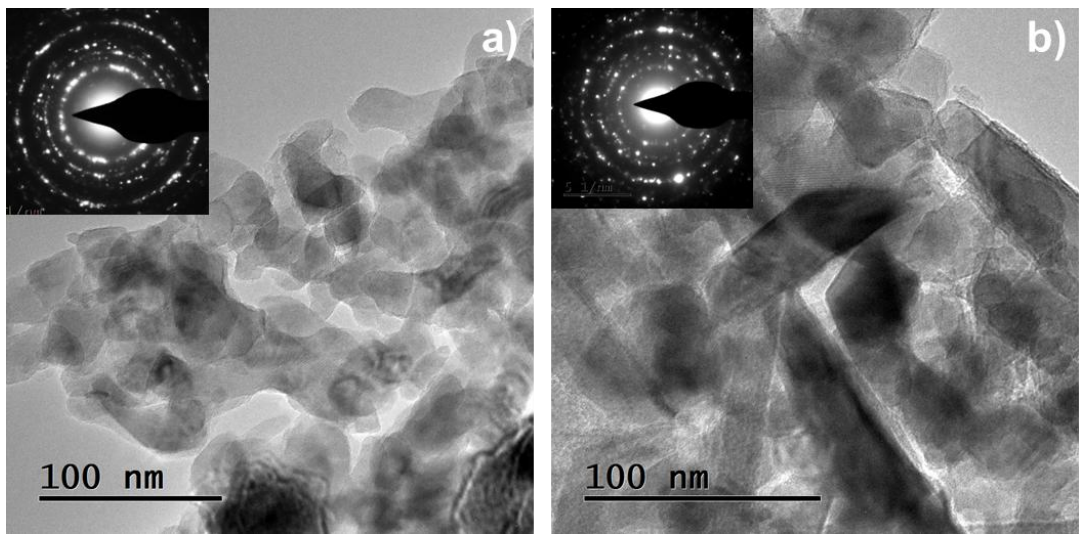

**Figure 8.** TEM micrographs and SAED of an RPAVD-$O_2$ film (a) before and (b) after annealing to induce the transformation to $VO_2$.

Grazing incidence X-ray diffraction (GIXRD) analysis was performed to evaluate the crystallinity of the samples. **Figure 9a)** displays the GIXRD diffractograms for a 450 nm aerogel-like $V_2O_5$ film and the corresponding $VO_2$ film after annealing. The diffractograms show that both types of films are crystalline, showing the transition from orthorhombic $V_2O_5$ to monoclinic $VO_2$ after the annealing, similar to what was observed by HRTEM. In both cases, the width of the XRD peaks can be attributed to the finite crystallite size, which influences the peak broadening according to the Scherrer equation. This broadening reflects the minimum crystallite size within the sample, as smaller crystallites produce wider peaks.[59] The as-deposited $V_2O_5$ aerogel-like sample exhibits well-defined peaks at lower angles between 15º and 35º. These peaks correspond to the planes (020), (001), (110), (031), and (130), characteristic of the orthorhombic structure of $\alpha$-$V_2O_5$ (ICSD No. 60767, *Pmmn*, a = 11.5 Å, b = 3.5 Å, c = 4.3 Å and $\alpha = \beta = \gamma = 90º$). According to the Scherrer equation, the crystallite size for the $V_2O_5$ aerogel is estimated to be 6.5 nm. This crystalline size is consistent with the HRTEM results shown in **Figure 5a)**, where small crystals of around 5 nm (note that the walls were as thin as 5-10 nm) were surrounded by an amorphous matrix. These results further demonstrate that the cycling plasma polymerization and etching can give rise to crystalline $V_2O_5$ thin films at low temperatures (i.e., 120 ºC during the etching process in each cycle) without requiring high temperatures to complete the film oxidation.

After annealing, the samples completely lose the diffraction peaks of the $V_2O_5$ and develop mainly four additional peaks that correspond to the (0-11), (21-1), (002), and (21-2) planes, indicative of the monoclinic structure of $VO_2$ (M1) (ICSD No. 74705, *P21/c*, a = 5.7 Å, b = 4.5 Å, c = 5.4 Å nm, $\alpha = \gamma = 90º$ and $\beta = 122.61º$). According to the Scherrer equation, the crystallite size for the $VO_2$ aerogel after annealing is estimated to be 6.6 nm, which is also in good agreement with the crystalline sizes observed by HRTEM (10-20 nm). Importantly, no evidence can be observed for other crystalline phases, such as $V_2O_5$, $V_6O_{13,}$ or $V_2O_3$, indicating the high purity of the films. These results constitute compelling evidence validating the optimization of the annealing process for $VO_2$ film production.

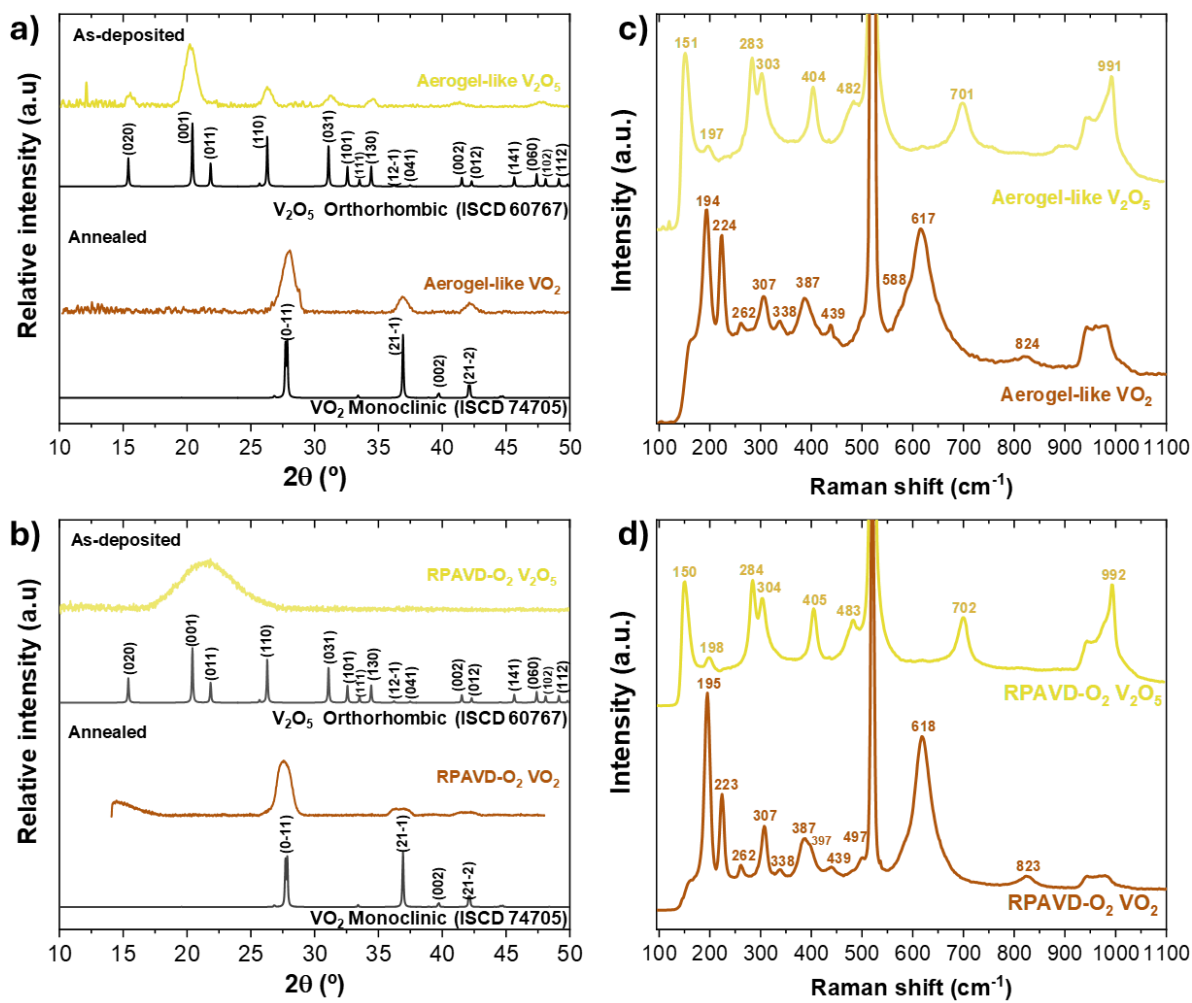

**Figure 9.** (a) GIXRD patterns of an as-deposited 450 nm thick $V_2O_5$ aerogel-like film and the same film after thermal annealing. (b) GIXRD of a 110 nm thick $V_2O_5$ film deposited by RPAVD-$O_2$ before and after annealing. The GIXRD references of $V_2O_5$ and $VO_2$ in (a) and (b) are included for comparison. (c) Raman spectra of a 450 nm thick aerogel-like $V_2O_5$ film before and after annealing. (d) Raman spectra of a 110 nm thick RPAVD-$O_2$ film before and after annealing. The peak at ~520 cm$^{-1}$ from the Si(100) substrate is visible in the Raman spectra.

**Figure 9b)** shows the GIXRD of a $VO_2$ sample prepared by RPAVD-$O_2$ of 110 nm after and before the annealing process. The reference $V_2O_5$ porous samples (**Figure 9b)**) do not show any diffraction peaks. After annealing, the resulting $VO_2$ film is crystalline, showing peaks similar to those of the aerogel-like $VO_2$ films corresponding to the monoclinic phase without the presence of additional peaks that could be ascribed to other vanadium oxide phases. The lack of diffraction patterns of the $V_2O_5$ films may be very likely ascribed to the low thickness of the analyzed samples because the HRTEM analysis presented in **Figure 8** showed that the sample is nanocrystalline.

The vanadium oxide films were characterized by room temperature Raman before and after annealing and analyzed according to the literature.[60–64] As depicted in **Figure 9c)**, the as-deposited aerogel-like thin film exhibits characteristic peaks indicative of pure crystalline α-$V_2O_5$, congruent with the previous XRD and TEM characterizations, with no observable traces of other vanadium oxides or compounds. The structural analysis reveals 21 Raman modes ($7A_g + 3B_{1g} + 7B_{2g} + 4B_{3g}$). Notably, the high-frequency Raman peak observed at 992 cm$^{-1}$ ($A_g$) corresponds to the terminal oxygen V=O stretching mode, resulting from unshared oxygen. Another significant peak at 702 cm$^{-1}$ ($B_{1g}$/ $B_{3g}$) is attributed to the asymmetric stretching mode of doubly coordinated oxygen in V-$O_{(2)}$-V, arising from oxygen shared at the corners common to two pyramids. Unfortunately, the 528 cm$^{-1}$ ($A_g$) and 507 cm$^{-1}$ peaks overlap with the silicon substrate peak. These peaks are associated with the triply coordinated oxygen (V-$O_{(3)}$-V) stretching mode, resulting from oxygens shared at the edges common to the three pyramids. Further insights into the vibrational characteristics include the identification of peaks at 483 ($A_g$) and 304 cm$^{-1}$ ($A_g$), corresponding to the bending vibrations of bridging V-O-V (doubly coordinated oxygen) and triply coordinated oxygen (V-$O_{(3)}$-V) bonds, respectively. Additionally, the peaks at 405 ($A_g$) and 284 cm$^{-1}$ ($B_{1g}$/ $B_{3g}$) are associated with the bending vibration of V=O bonds. Two low-frequency Raman peaks, discernible at 198 ($A_g$/ $B_{2g}$) and 150 cm$^{-1}$ ($B_{1g}$/ $B_{3g}$), are indicative of lattice vibrations and provide insights into the vibrational modes associated with the layered structure and the structural features of α-$V_2O_5$.[60,65–67]

The Raman spectrum for the aerogel-like annealed sample displays all the peaks attributed to $VO_2$ (M1), consistent with previously reported data[60,61]. According to group theory, the M1 phase of $VO_2$ possesses 18 Raman active modes ($9A_g$ and $9B_g$), with 12 modes observed in the $VO_2$ thin film at ~138, 194, 223, 262, 307, 338, 387, 439, 442, 497, 583, and 613 cm$^{-1}$ (**Figure 9c**). Importantly, no other phases or polymorphs were detected. The slight discrepancies between our data and those in the literature refer to small shifts in the Raman peak positions (in cm$^{-1}$). These shifts can be attributed to the pronounced difference in thermal expansion coefficients between $VO_2$ and the silicon substrate, which induces heightened internal strain in films deposited on this substrate, thus explaining the observed variations.[62] The low-frequency phonons at 194 and 223 cm$^{-1}$ are associated with lattice motion involving V-V bonds, while the remaining peaks correspond to vibrational modes of V-O bonds. The bands in the low wavenumber region (< 400 cm$^{-1}$) are attributed to V−O−V bending modes. As the wavenumber transitions to the intermediate range (400−800 cm$^{-1}$), the bands correspond to V−O−V stretching modes. The broad peak at 613 cm$^{-1}$ is a convolution of the 588, 613 and 661 cm$^{-1}$ peaks. The bands in the high wavenumber range (> 800 cm$^{-1}$) can be assigned to V=O stretching modes indicative of distorted octahedra and square pyramids.

The interpretation of the Raman spectra of both RPAVD-$O_2$ as-deposited and annealed are similar, presenting Raman bands corresponding to crystalline $V_2O_5$ and $VO_2$, respectively (**Figure 9d**). Thus, it is noticeable that there is a discernible level of crystallinity in the as-deposited porous sample that the GIXRD analysis (**Figure 9b**) could not capture. This effect might be related to the low thickness and the random orientation of the nanocrystallites with few atomic planes for diffraction.

The surface composition of the as-deposited and annealed vanadium oxide aerogel-like films was studied by XPS and analyzed according to the literature.[47,63] **Figure 10a)** shows the core-level V2p of an aerogel film before and after annealing. The as-deposited aerogel-like film presents exclusively $V^{5+}$ bands at ~517.6 eV ($V^{5+}$, $2p_{3/2}$) and ~524.9 eV ($V^{5+}$, $2p_{1/2}$) without traces of other vanadium species. This result indicates that the remote oxygen plasma treatment effectively oxidized the vanadium species (mainly $V^{4+}$) from the VOPc sacrificial plasma polymer to their highest oxidation state. The $VO_2$ aerogel-like sample obtained after the annealing treatment shows two majority $V^{4+}$ core level bands at ~516.4 eV ($V^{4+}$, $2p_{3/2}$) and ~523.7 eV ($V^{4+}$, $2p_{1/2}$) in the same binding energies and two additional less intense bands at the $V^{5+}$ binding energies. Considering the results of the previous GIXRD, HRTEM, and Raman characterizations, the small percentage of $V^{5+}$ at the surface of the annealed films is likely due to surface oxidation. The RPAVD-$O_2$ films present surface compositions similar to those of the aerogel-like films after and before the annealing.

Additional details about the XPS analyses of the RPAVD-Ar VOPc sacrificial plasma polymers and sublimated samples are included in **S2.**

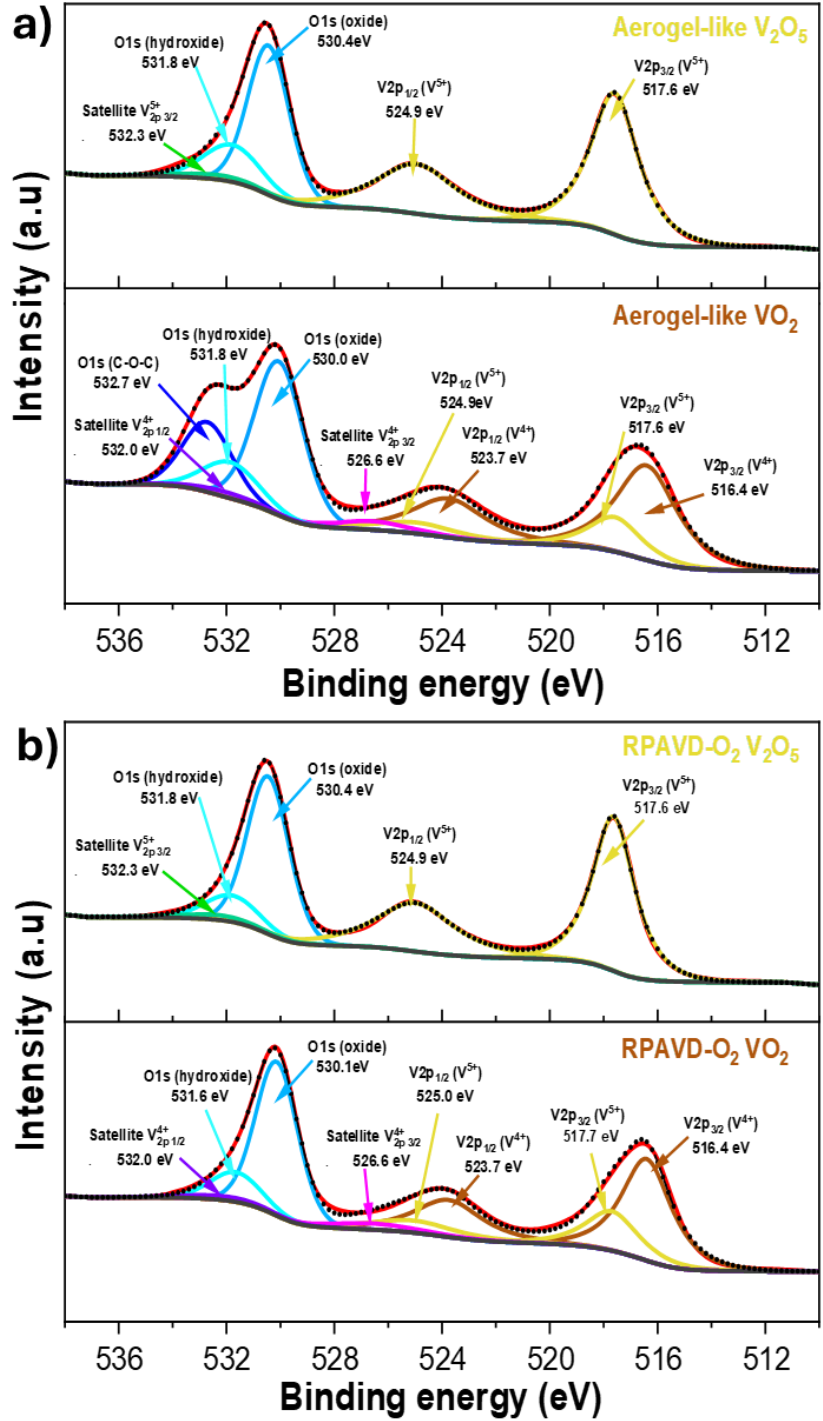

**Figure 10.** V2p+O1s XPS spectra for (a) aerogel-like and (b) RPAVD-O$_2$ films before and after the annealing process.

## 3.4 Optical performance

The thermochromic performance of the samples was studied by UV-VIS-NIR transmittance spectroscopy at controlled temperature for samples deposited on fused silica.

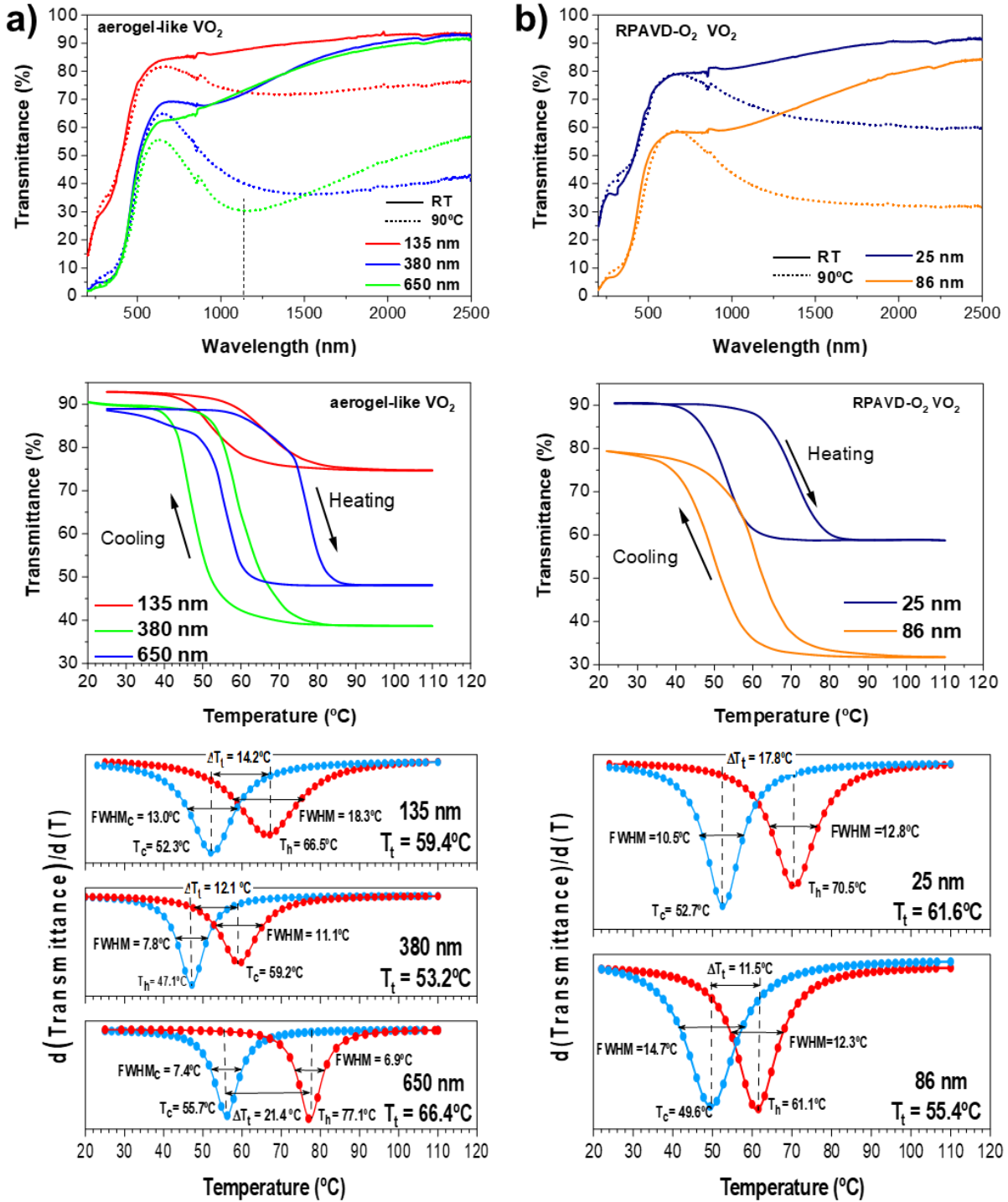

**Figure 11.** Temperature-dependent transmittance spectra in the UV-VIS-NIR region at 25 °C and 90 °C (top), measured hysteresis loops (middle), and first derivative $(d(T_f)/d(T))$ versus temperature (bottom) for a set of aerogel-like VO$_2$ thin films (a) and RPAVD-O$_2$ VO$_2$ thin films (b).

**Figure 11a)** top shows the results of three aerogel-like samples measured at room temperature (25 °C) and 90 °C. The overall shape of the spectra is characteristic of VO$_2$.[68,69] The films exhibit strong UV absorption and visible transparency at room temperature, reaching transmittance values of 85-63% at 630 nm, depending on the layer thickness. Besides, the aerogel-like sample presents a high transparency in the NIR region, with transmittance values at $\lambda$=2000 nm in the 93-89% range, depending on the layer thickness. All the samples have transmitance values higher than 90% at 2500 nm. At 90 °C, the aerogel-like VO$_2$ samples show a significant reduction of the IR transmittance due to the transition from the insulating to the metallic phase, reaching values at $\lambda$=2000 nm in the 74.4-38.4% range, depending on the thickness.

For the assessment of the films' potential for smart window application, we have calculated their key thermochromic parameters. These include luminous transmittance (T$_{lum}$), solar modulation ($\Delta$T$_{sol}$), and near-infrared modulation ($\Delta$T$_{IR}$), all derived from the transmittance values at each thickness. These properties are quantified according to the following expression:

$$T_i = \frac{\int T(\lambda)\phi_i(\lambda)d\lambda}{\int \phi_i d\lambda}$$

and $\Delta T_i = T_{i,cold}$-$T_{i,hot}$. $T(\lambda)$ represents the transmittance value for each wavelength; $i$ denotes *lum*, *sol*, or *IR*; $\varphi_{lum}(\lambda)$ is the standard photopic efficiency of vision (spectral sensitivity of the human eye) in the range 380-780 nm, from values tabulated by the International Commission on Illumination.[70] $\varphi_{sol}(\lambda)$ and $\varphi_{IR}(\lambda)$ are the solar irradiance spectrum for an air mass of 1.5 when the sun is at 37º above the horizon in the 280-2500 nm range and 780-2500 nm, respectively, with values tabulated by the American Society for Testing and Materials[59]. Cold and hot values correspond to the curves measured at 25 ºC and 90 ºC after and before the thermochromic transition. All these results are gathered in **Table 1**.

**Table 1.** Optical thermochromic parameters during the phase transition of selected aerogel-like and RPAVD-O$_2$ samples. Low and high T correspond to 25 and 90 ºC, respectively.

| Sample | Thickness (nm) | $T_t$(ºC) | $T_{lum}$ (%) | | | $T_{sol}$ (%) | | | $T_{IR}$ (%) | | |
|---|---|---|---|---|---|---|---|---|---|---|---|
| | | | Low T | High T | $\Delta T_{lum}$ | Low T | High T | $\Delta T_{sol}$ | Low T | High T | $\Delta T_{IR}$ |
| Aerogel-like | 135 | 59.4 | 79.6 | 77.6 | 2.0 | 80.2 | 73.3 | 6.9 | 87.9 | 74.9 | 13.0 |
| | 190 | 58.5 | 77.3 | 74.7 | 2.6 | 78.0 | 70.8 | 7.2 | 85.8 | 72.6 | 13.3 |
| | 380 | 53.2 | 58.6 | 55.8 | 2.8 | 61.2 | 46.5 | 14.7 | 74.3 | 44.5 | 29.8 |
| | 650 | 66.4 | 54.4 | 49.2 | 5.2 | 58.6 | 39.8 | 18.8 | 73.4 | 37.9 | 35.5 |
| RPAVD-O$_2$ | 25 | 61.6 | 74.0 | 74.0 | 0.0 | 75.4 | 69.1 | 6.2 | 83.1 | 69.0 | 14.2 |
| | 85 | 55.4 | 54.0 | 52.4 | 2.4 | 54.8 | 44.5 | 10.3 | 63.8 | 42.2 | 21.6 |

One intriguing aspect of the aerogel-like thermochromic films is their departure from the typical tens of nanometer thickness range of VO$_2$ thermochromic compact films due to the inherent opacity of this material.[68,71] As seen in **Figure 11a)**, the thickness range of aerogel-like films is up to several hundreds of nanometers. The first notable behavior concerning $T_{lum}$ is observed across all the aerogel-like samples. Generally, in low-thickness thermochromic films, the $T_{lum}$ value at low temperatures is consistently higher than at high temperatures. Various studies have demonstrated that $VO_2$ thermochromic films have a threshold thickness value beyond which the $T_{lum}$ at high temperatures surpasses the $T_{lum}$ at low temperatures. This reversal in $\Delta T_{lum}$ is attributed to interference effects[68,72,73] and adversely affects $\Delta T_{sol}$, decreasing its value when the thickness exceeds the mentioned threshold. However, this phenomenon is not manifested in the aerogel-like samples, suggesting that the threshold thickness for the reversal in $\Delta T_{lum}$ has not yet been reached (see **Figure 11a)** top and **Table 1**). Thus, for aerogel-type samples, the $\Delta T_{sol}$ values for the thinnest films are 6.9% for a thickness of 135 nm, increasing by only 0.3% (i.e., to 7.2%) as the film thickness increases to 190 nm. Meanwhile, $T_{lum}$ remained almost unchanged (79.6/77.6% and 77.3/74.7%, respectively, for low/high T). However, when the thickness doubles to 380 nm, the $\Delta T_{sol}$ value more than doubles, from 7.2% to 14.7%, with a $T_{lum}$ value of 58.6/55.8% (low/high T). Further increasing the film thickness to 650 nm, the $\Delta T_{sol}$ value reaches a remarkable 18.7%, accompanied by a mild decrease in $T_{lum}$ to just 54.4/49.2%. These findings demonstrate that aerogel-type samples exhibit an optimum compromise for balancing $T_{lum}$ and $\Delta T_{sol}$ through porosity/thickness, positioning the aerogel-like samples among the most outstanding thermochromic films reported.[4,9,11,23,74]

Another notable feature of the aerogel-like thermochromic films is their remarkably high $\Delta T_{IR}$ values, which are substantially higher than those observed in traditional $VO_2$ thermochromic films.[17,18] For the 135 nm and 190 nm samples, the $\Delta T_{IR}$ values are 13% and 13.3%, respectively. However, with increasing thickness to 380 nm, $\Delta T_{IR}$ rises considerably to 29.8% and reaches a maximum of 35.5% for the 650 nm films. This remarkable infrared modulation further underscores the excellent performance of aerogel-like $VO_2$ films, making them highly promising for smart window applications where both visible light and infrared modulation are crucial.

The semiconductor-to-metal switching characteristics of the thin films were examined by recording the temperature-dependent optical transmittance during both the heating and cooling processes at λ=2000 nm, as illustrated in the central panel of **Figure 11**. To calculate the transition temperature ($T_t$), the derivative of the optical transmittance ($T_r$, to distinguish from temperature T) with respect to temperature (T) was plotted against the temperature (T) for each heating and cooling branch, as shown in the bottom panel of **Figure 11**. This graphical representation of the derivative allows for identifying two inflexion points indicating the phase transition in the optical properties of the films. Each curve was fitted to a Lorentzian function, as shown in the bottom panel of **Figure 11**. $T_h$ and $T_c$ are the transition temperatures for the heating and cooling cycles, while the average transition temperature ($T_t$) is then calculated as $T_t = (T_h + T_c)/2$.

The full width at half maximum (FWHM) values (bottom panel in **Figure 11**) provide an insight into the sharpness of the transition, i.e., how abrupt the metal-insulator transition is in the thin film during the heating ($FWHM_h$) and cooling ($FWHM_c$) phases. The hysteresis width ($\Delta T_t$) is defined as the difference between $T_h$ and $T_c$, providing a measure of the hysteresis of the process. A wide hysteresis width indicates that the material can maintain its metallic or insulating phase over a broader range of temperatures during cooling or heating. In contrast, a smaller hysteresis width suggests greater symmetry in the transition.

The interpretation of the transition temperatures and hysteresis loops of the aerogel-like films as a function of the thickness is not straightforward. The average transition temperature (**Table 1**) decreases with the thickness, starting at 59.4 °C for the 135 nm thick sample and decreasing to 58.5 and 53.2 °C for the 190nm and 380 nm thick samples, respectively. This latter sample presents the minimum $T_t$ for the studied samples, which is significantly lower than the $T_t$ corresponding to crystalline undoped $VO_2$ (Tt ~ 68 °C).[68,73] The transition hysteresis also decreases from $\Delta T_t$ =

14.2 ºC to $\Delta T_t = 12.1$ ºC in the same thickness range (bottom panel in **Figure 12a)**). For these samples, the reduced transition temperatures and the hysteresis evolution can be attributed to factors such as the distribution and size of $VO_2$ crystals, the effect of the strain of the $VO_2$ nanocrystal lattices observed in the HRTEM analysis, and the impact of the increasing porosity of the aerogel-like films with the thickness.[69,75]

However, for the 650 nm thick film, $T_t$ increases to 66.4 ºC, a value close to the $VO_2$ nominal $T_t$, and the transition hysteresis increases to 21.4 ºC (**Figure 11a)**). These results indicate that the $VO_2$ crystal structure is relaxed in this sample. The evolution of the density of the aerogel-like $VO_2$ films with thickness shows a density decrease with thickness up to 380 nm and then a slight increase for the film of 650 nm thick, giving rise to the plateau in **Figure 1d)**. A similar plateau can be observed in the $V_2O_5$ precursor films (**Figure 1c)**) and also in partly crystalline $TiO_2$ films prepared by the same methodology.[34] The shape of the curve in **Figure 1d)** points out a thickness gradient density that decreases with the thickness. Plotting the section density of every cycle as shown in **Figure 12**, it can be noted that the minimum density is observed at 380 nm, and then it increases slightly for the 650 nm thick sample. This thickness increase is likely linked to the percolation of $VO_2$ crystals, leading to the observed rise in $T_t$. Thus, several authors have reported $T_t$ close to the nominal value of $VO_2$ in percolated nanoporous structures[76] and in $VO_2$ crystalline nanoparticles.[32,71]

Another significant difference that can be observed in the shape of the 650 nm thickness aerogel-like sample's optical transmittance curve is the presence of an absorption band at ca. 1150 nm in the high-temperature spectrum (top panel in **Figure 11a)**), highlighted with a vertical dashed line).

This absorption corresponds to the formation in the films of a localized surface plasmon resonance (LSPR), which is not observed for lower thicknesses where light transmission decreases monotonically toward long wavelengths. Surface plasmon resonances in the range 1100-1200 nm have been reported for VO$_2$ nanoparticles of different shapes and aspect ratios[32,71,77] and recently in percolated nanoporous structures.[76] The formation of a plasmonic structure in the 650 nm film is congruent with the percolation of VO$_2$ nanocrystals to give rise to larger structures with an associated increase in film density, as shown in **Figure 12**. Notably, VO$_2$ localised surface plasmon resonances have been proposed as an effective tool to improve the solar regulation efficiency of VO$_2$-based thermochromic systems, which can absorb infrared radiation without affecting visible light absorption.[32,68,76].

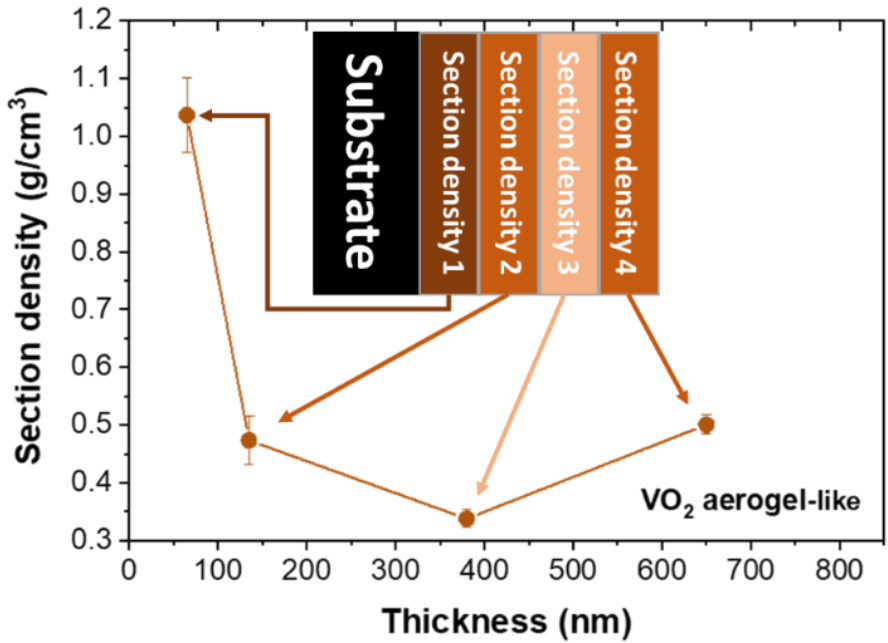

**Figure 12.** Evolution of the section density of the film as a function of the thickness for the VO$_2$ aerogel-like films calculated using the data from **Figure 1d)**.

On the other hand, the thermochromic performance (**Figure 11b)** and **Table 1)** of the RPAVD-$O_2$ $VO_2$ films are different from those of the aerogel-like samples. In these samples, the thermochromic operation is restricted to ultrathin films with thicknesses lower than 100 nm, similar to $VO_2$ films prepared by magnetron sputtering, pulsed laser deposition, and other techniques.[74,75,78,79] Thus, increasing the thickness, $\Delta T_{sol}$ rises by 4.1%, while $T_{lum}$ decreases drastically by 20%. The $\Delta T_{IR}$ values are notably high for the two samples, reaching 21.6% for the thicker sample. The Tt decreases with the thickness from a value close to the nominal value of $VO_2$ to 55.4 °C, a value significantly lower and in the range of the aerogel-like films with thicknesses lower than 380 nm. These findings are likely related to the transition from continuous percolated porous films to higher porosity films formed by more isolated nanograins observed in **Figure 3b)**.

### 3.5 Encapsulation of aerogel-like films for environmental protection

$VO_2$ is thermodynamically unstable and can be oxidised to $V_2O_5$ under air exposure over several months and accelerated under humid environments.[80] This environmental instability limits the practical applicability of thermochromic $VO_2$ films, nanoparticles, and nanocomposites, mainly for smart window applications[81]. Currently, research is underway to synthesise hybrid materials that can almost entirely prevent the oxidation of $VO_2$ without compromising its thermochromic properties.[9,11,81,82] Thus, different solutions have been proposed to effectively delay the oxidation process, such as the use of inorganic protective coatings and shells as $SiO_2$[9], $Al_2O_3$[83], $ZnO$[84], and polymers as PMMA.[85]

In this section, we present a preliminary study on the encapsulation of aerogel films using a high-performance polymer synthesized with the same reactor and technique used to synthesize the sacrificial vanadium polymer films. This encapsulating polymer is deposited by RPAVD of adamantane (hereafter ADA films). These ADA films have been applied to encapsulate hybrid halide perovskite cells, enabling their use in high humidity conditions and underwater operation[86] and MoS$_2$ 2D layers for strain engineering.[41] The ADA films are fully transparent in the visible and IR regions with a refractive index of 1.57 at 550 nm. Besides, the films are insoluble, thermally stable, and robust against harsh chemical environments.[38] Another important aspect is the possibility of providing mechanical stability to the aerogel-like film surfaces thanks to the relatively high hardness and high Young's modulus values of the films (7.5-10.0 GPa).[41] Note that this aspect is crucial for using aerogel-like films in smart-coating applications.

**Figure 13** shows the spectral transmittance of a 375 nm aerogel-like film before (purple line) and after (red line) the encapsulation with a 150 nm thick ADA film (both at RT and 90ºC). In addition, the ADA encapsulated sample has been measured after a long period of exposure to atmospheric conditions (humidity values between 35-50% for 550 days). The transmittance of the aerogel-like thin film right after the encapsulation with the adamantane plasma polymer remains practically unchanged (**Figure 13a)**). Only minimal changes in the NIR transmittance are appreciated, of less than 5% compared to the original film, likely due to the anti-reflectance effects of the adamantane film conformal coating. However, the thermochromic parameters are affected. As seen in **Figure 13b)**, the hysteresis of the ADA-VO$_2$ film maintains practically the same shape and symmetry but is shifted to higher temperature values. The parameters are summarised in **Table 2**. It can be noted that encapsulation with the adamantane film increases the transition temperature by 12.9°C compared to the original film (from 53.2 to 66.1°C). However, the values of T$_{lum}$ remain imperceptible, decreasing from 58.6% to 56.5%, while the values of $\Delta$T$_{sol}$ increase from 14.7% to 16.5%.

**Table 2.** Optical parameters of VO$_2$ aerogel-like sample coated with 150 nm of RPAVD-ADA coating as deposited and after 550 days of ambient exposure.

| Sample | T$_t$ (ºC) | T$_{lum}$ (%) | | | T$_{sol}$ (%) | | | T$_{IR}$ (%) | | |
|---|---|---|---|---|---|---|---|---|---|---|
| | | Low T. | High T. | ΔT$_{lum}$ | Low T | High T. | ΔT$_{sol}$ | Low T. | High T. | ΔT$_{IR}$ |
| VO$_2$ aerogel as annealed | 53.2 | 58.6 | 55.8 | 2.8 | 61.2 | 46.5 | 14.7 | 74.3 | 44.5 | 29.8 |
| ADA/VO$_2$ Day 0 | 66.1 | 56.5 | 52.4 | 4.1 | 61.4 | 44.8 | 16.5 | 77.2 | 45.0 | 32.2 |
| ADA/VO$_2$ Day 550 | 67.8 | 56.5 | 53.6 | 2.9 | 59.7 | 47.1 | 12.6 | 73.6 | 48.6 | 25.0 |

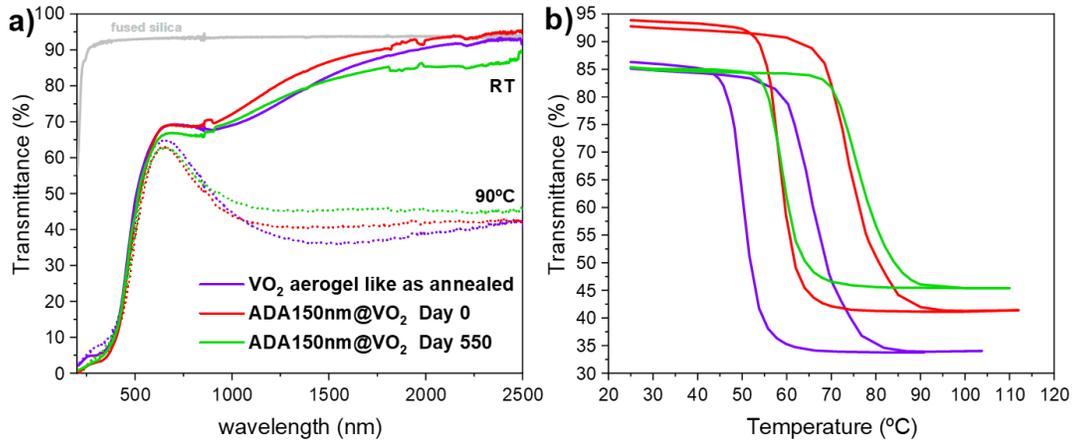

**Figure 13.** Effects of 150 nm adamantane plasma polymer encapsulation on a 380 nm aerogel-like VO$_2$. (a) Temperature-dependent optical transmittance spectra in the UV-VIS-NIR region at room temperature (solid lines) and 90 ºC (dotted lines). (b) Hysteresis loops corresponding to the films in (a).

The encapsulating effect of the adamantane film becomes evident in the transmittance spectra of the hot and cold states after 550 days of air exposure. In this case, the ΔT$_{IR}$ compared to the measurements taken on day 0 decreases only by 7.2%, whereas the hysteresis of the process remains practically unchanged, and $T_t$ increases very slightly by 1.7 °C. In addition, $\Delta T_{sol}$, decreases by 4.0% (from 16.5 to 12.6%), while $T_{lum}$ remains almost unaffected.

The above results show the potential of the remote plasma polymerization technique to encapsulate aerogel-like $VO_2$ films and significantly increase their environmental and mechanical stabilities. This study is positive but preliminary. A more comprehensive analysis is needed to optimize parameters such as the optimal thickness of the polymeric layer, degree of crosslinking as a function of the preparation conditions, and conformality of the polymer to the internal porous structure. Systematic characterization of the encapsulated $VO_2$ films as a function of temperature and degree of humidity is also needed. These studies are currently underway.

## 4. Conclusions

In a previous work, we presented a methodology for fabricating aerogel-like titanium oxide films for photonic applications, such as anti-reflective coatings and ultraporous anatase photoelectrodes for perovskite solar cells, using a low-temperature process based on remote plasma deposition and processing.[34] In this work, we demonstrate that this technique can be directly applied to developing $VO_2$ aerogel-type thermochromic films by first synthesizing $V_2O_5$ aerogel-like films, followed by a reductive annealing step. These results confirm the generality of the synthetic method and its potential to develop ultraporous oxide films with a wide range of compositions. In this procedure, high porosity is achieved by remote plasma polymerization of a high C to V ratio precursor, such as vanadyl phthalocyanine with C/V= 32, and subsequent plasma etching removal of the organic part of the film (C and N species) by formation of volatile compounds which are pumped out. Simultaneously, oxidation of the $V^{4+}$ cation occurs, giving rise to partially crystalline $V_2O_5$ films. Utilizing a cyclic process of polymerization and oxidation, we have been able to control the thickness and porous structure of the material. After annealing under Ar atmosphere, we obtained ultraporous films formed by nanocrystalline monoclinic $VO_2$ without amorphous or any additional crystalline oxide phases. These $VO_2$ aerogel films are thermochromic, exhibiting very high luminous transmittances in the range of 79.6%-54.4% and solar modulations as high as 14.7% and 18.8% for films of 380 and 650 nm, respectively. The thicker films develop a surface plasmonic resonance at 1150 nm for temperatures higher than $T_t$, which we attribute to the percolation of the nanocrystalline structure. These plasmonic structures contribute to the enhancement of thermochromic modulation efficiency.

We have fabricated more compact $V_2O_5$ films using the same precursor and an oxygen plasma RPAVD process in a single synthetic step. By applying the same reductive annealing treatment used for the aerogel films, we obtained crystalline thermochromic $VO_2$ films showing light transmission values above 50% and solar modulation up to 10.3% in a thickness range below 100 nm. These results prove the versatility of the synthetic approach that can be applied to synthesizing compact and ultraporous $VO_2$ films from the same precursor by changing the synthesis conditions. The main objective of this work has been to show the potentiality of the plasma process to generate novel materials such as low-temperature conformal aerogel-like thermochromic $VO_2$ films. However, due to their general character, the methodology is also compatible with doping with elements (i.e., W, Mg) and multilayered optical designs that are extensively reported for improving the optical transmission and thermochromic performance of $VO_2$ films for smart solar modulation. Besides, the synthetic methodology also offers a practical option for encapsulating ultraporous structures with robust, stable, and completely transparent coatings. Another essential aspect to stress is the industrial scalability and the environmental compatibility characteristic of the plasma processes.

**Associated content**

**Supporting information** Optical, SEM, and XPS characterizations of VOPC RPAVD films

ACKNOWLEDGMENTS

We thank the projects PID2022-143120OB-I00, TED2021-130916B-I00, PCI2024-153451, and PID2021-123879OB-C21 funded by MCIN/AEI/10.13039/501100011033 and by "ERDF (FEDER) a way of making Europe, the Fondos Nextgeneration EU, and Plan de Recuperación, Transformación y Resiliencia." Project ANGSTROM was selected in the Joint Transnational Call 2023 of M-ERA.NET 3, an EU-funded network of about 49 funding organizations (Horizon 2020 grant agreement No 958174). We also thank the Consejería de Economía, Conocimiento, Empresas y Universidad de la Junta de Andalucía (PAIDI-2020 through projects P18-RT-3480). The project leading to this article has also received funding from the EU H2020 program under grant agreement 851929 (ERC Starting Grant 3DScavengers).

**Supporting Information**

**Enhanced Luminous Transmission and Solar Modulation in Thermochromic VO₂ Aerogel-Like Films via Remote Plasma Deposition**

*Jose M. Obrero-Pérez,[a] Gloria Moreno-Martínez,[a] Teresa C. Rojas,[b] Francisco J. Ferrer,[c,d] Francisco G. Moscoso,[e] Lidia Contreras-Bernal,[a,f] Javier Castillo-Seoane,[a] Fernando Núñez-Gálvez,[a] Francisco J. Aparicio,[a] Ana Borras,[a] Juan R. Sánchez-Valencia,[a*] Angel Barranco[a*]*

a) Nanotechnology on Surfaces and Plasma Laboratory, Materials Science Institute of Seville (CSIC-US), C/ Américo Vespucio 49, 41092, Seville, Spain.

b) Tribology and Surface Protection Group. Materials Science Institute of Seville (CSIC-US), C/ Américo Vespucio 49, 41092, Seville, Spain.

c) Centro Nacional de Aceleradores (Universidad de Sevilla, CSIC, Junta de Andalucía). Avda. Tomas Alba Edison 7, 4092, Sevilla

d) Departamento de Física Atómica, Molecular y Nuclear, Universidad de Sevilla, Aptdo 1065, 41012 Sevilla, Spain

e) Center for Nanoscience and Sustainable Technologies (CNATS), Departamento de Sistemas Físicos, Químicos y Naturales, Universidad Pablo de Olavide, Ctra. Utrera km. 1, Sevilla 41013, Spain.

f) Departamento de Ingeniería y Ciencia de los Materiales y el Transporte, EPS-Universidad de Sevilla, c/Virgen de África 7, 41011, Sevilla, Spain.

Corresponding author e-mail: angel.barranco@csic.es; jrsanchez@icmse.csic.es

**S1.–UV-Vis spectra and FESEM micrographs of RPAVD-Ar polymer films**

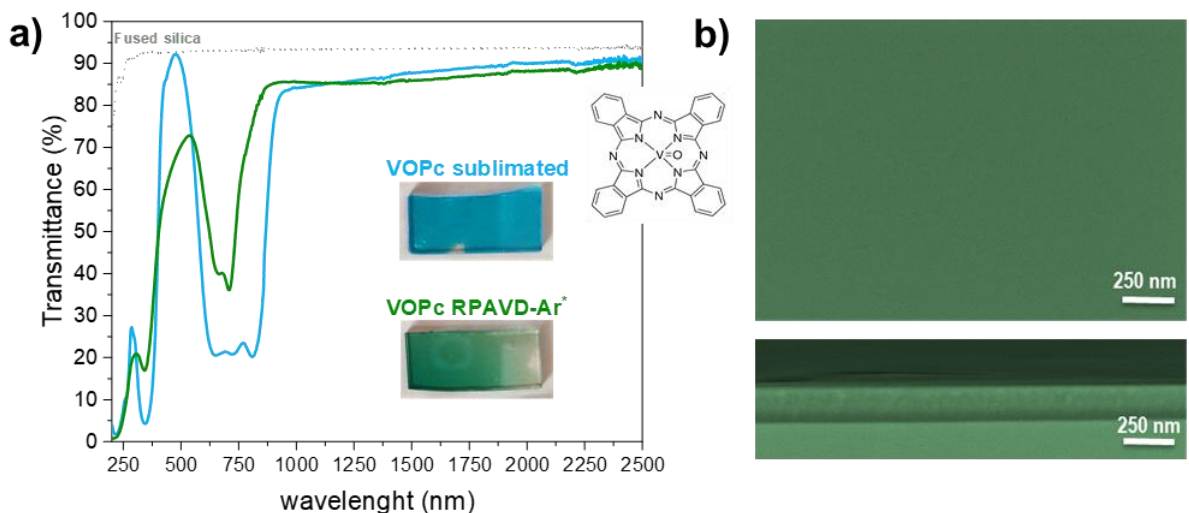

**Figure S1.** (a) UV-VIS-NIR Transmittance spectra of a plasma polymerized VOPc film by RPAVD-Ar and a reference sublimated VOPc films. It also includes photographs of the films deposited on fused silica. (b) FESEM normal and cross-sectional micrographs of the RPAVD-Ar VOPc layer used as a sacrificial layer to synthesize aerogel-like $V_2O_5$ and $VO_2$ films.

**S2. XPS characterisation of sublimated and RPAVD-Ar VOPc films.**

The sublimated and VOPc plasma polymeric films deposited by RPAVD-Ar were analysed by XPS. The survey spectra and the atomic/relative percentages of each element (V, C, O, and N) are presented in **Figures S2a)** and **S2b)**, respectively. Significant differences were observed in the relative amounts of N, C, and O obtained from XPS analysis with respect to V (**Figure S2b**). Specifically, $(N/V)_{XPS}$=5.0, $(C/V)_{XPS}$=27.0, and $(O/V)_{XPS}$=6.5, compared to the sublimated film, where $(N/V)_{sublimated}$=7.0, $(C/V)_{sublimated}$=31.0, and $(O/V)_{sublimated}$=1.0, values that closely approximate those of the empirical formula of VOPc ($C_{32}H_{16}N_8OV$). Polymerisation via RPAVD-Ar resulted in a slight decrease in C and N content, accompanied by an increase in oxygen content, which, in turn, led to the partial oxidation of the VOPc molecule and the generation of volatile carbon and nitrogen products. This oxygen enrichment is a common phenomenon in RPAVD techniques, attributed to post-deposition reactions with air and/or direct incorporation of residual oxygen from the reactor during deposition.[1,2]

To observe the changes undergone by the molecule during polymerization via RPAVD, XPS zone spectroscopy measurements were performed in the O1s+V2p, C1s, and N1s regions, and are shown in **Figures S2c), S2d) and S2e)**, respectively, for the sublimated and polymerized VOPc films. According to the O1s+V2p zone spectra (**Figure S2c)**), the oxidation state of vanadium in both cases can be estimated from the position of the V2p$_{3/2}$ peak. In the case of the sublimated VOPc film, a single intense peak at 416.4 eV is observed, confirming the exclusive presence of vanadium in its V$^{4+}$ form. However, when the molecule undergoes plasma polymerization, the residual oxygen in the chamber oxidizes some of the V$^{4+}$ to V$^{5+}$, as evidenced by the appearance of a second peak at 517.6 eV. Approximately 25% of the total V content oxidizes to the highest oxidation state during this process. The C1s spectra for the same samples are shown in **Figure S1d)**. Both exhibit C1s species characteristic of phthalocyanines: one corresponding to saturated C-H carbons, situated at a binding energy of 285.0 eV, and another in the unsaturated C=C carbon region at 284.5 eV. A second peak of C-N at 286.5 eV, associated with the carbon bonded to the pyridinic nitrogen of phthalocyanine (N$_\beta$), and a third peak of C=N at 288.0 eV, related to the carbon bonded to the pyrrolic nitrogen (N$_\alpha$), are highlighted. Additionally, a minor peak at 291.0 eV corresponds to the π-π* interaction of the phthalocyanine rings.

The incorporation of residual oxygen into the polymerized film is manifested by an increase in the peak at 288.0 eV, from 8.3% of the total C in the sublimated films to 15.2% in the polymerized films, where C=O bands are also observed. Thus, it is evident that some of the carbons located in the isoindole groups are partially oxidized. This phenomenon is more pronounced in the N1s spectra of **Figure S2e)**: in the sublimated film, only N$_\beta$ and N$_\alpha$ peaks at 398.8 and 399.3 eV, respectively, are observed in the same proportion. However, 80% of the surface N$\alpha$ peaks are partially fragmented during the plasma polymerization, generating more oxidized species, as indicated by a 400.2 eV peak assigned to the -N-(C=O)-O- species.

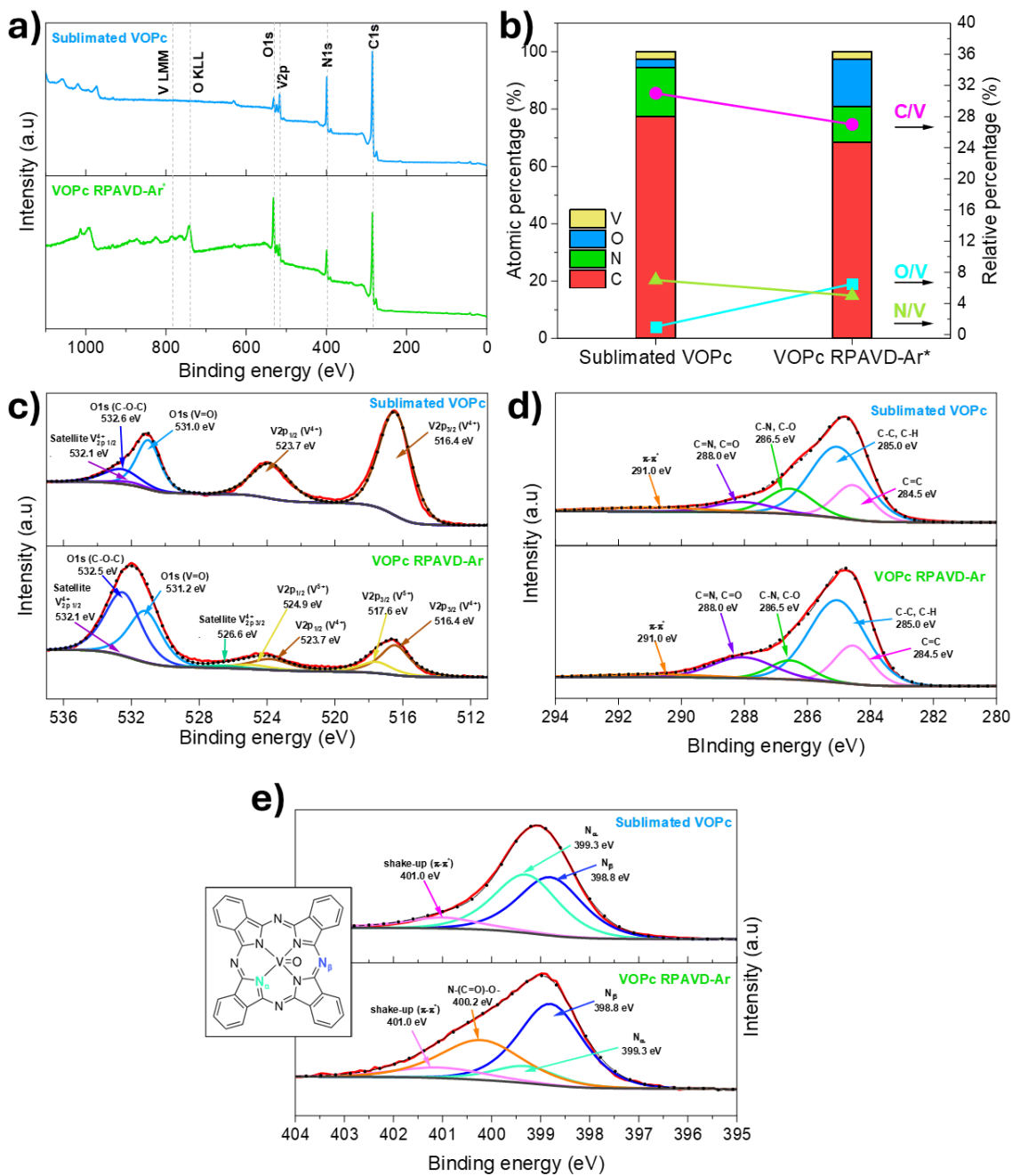

**Figure S2.** (a) XPS survey spectra of a reference VOPc sublimated films and a VOPc plasma polymer deposited by RPAVD-Ar. (b) Surface atomic percentages of the films determined by XPS. V2p+O1s (c), C1s (d) and N1s XPS spectra of the films in (a).